\documentclass[a4paper,11pt]{article}
\pdfoutput=1 
\usepackage{jheppub} 
\usepackage[T1]{fontenc} 

\usepackage{mathptmx}        
\usepackage{amsmath}
\usepackage{amssymb}
\usepackage{graphicx}        

\usepackage{color}
\usepackage{helvet}          
\usepackage{courier}         
\usepackage{dirtree}

\usepackage{subfig}

\usepackage{multicol}        
\usepackage[bottom]{footmisc}

\usepackage{hyperref}        
\hypersetup{colorlinks=true,urlcolor=blue}

\usepackage[misc]{ifsym}

\def\planck{{Planck}}

\def\actdr{{\sc ACT DR4}}

\def\be{\begin{equation}}
\def\ee{\end{equation}}
\def\ba{\begin{eqnarray}}
\def\ea{\end{eqnarray}}

\def\rd{r_{\rm d}}

\def\om{\Omega_{\rm m}h^2}
\def\Om{\Omega_{\rm m}}

\newcommand{\mksym}[1]{\ifmmode {\rm #1}\else #1\fi}

\newcommand{\TT}{\mksym{TT}}
\newcommand{\TE}{\mksym{TE}}
\newcommand{\EE}{\mksym{EE}}

\providecommand{\omm}{\omega_{\mathrm{m}}}

\newcommand{\lcdm}{\texorpdfstring{{$\rm{\Lambda CDM}$}}{ΛCDM}}

\newcommand{\LCDM}{\lcdm}
\newcommand{\lcdmb}{\texorpdfstring{{${\rm\Lambda CDM+}{\it b}$ }}{ΛCDM}}

\def\lsim{{}^<_{\sim}}

\title{\boldmath Primordial magnetic fields and the Hubble tension}
\author[a]{Karsten Jedamzik}
\author[b]{Levon Pogosian}

\affiliation[a]{Laboratoire de Univers et Particules de Montpellier, UMR5299-CNRS, Universite de Montpellier, 34095 Montpellier, France}
\affiliation[b]{Department of Physics, Simon Fraser University, Burnaby, BC, V5A 1S6, Canada}

\emailAdd{karsten.jedamzik@umontpellier.fr}
\emailAdd{levon@sfu.ca}
\abstract{Magnetic fields appear to be present in essentially all astrophysical environments, including galaxies, clusters of galaxies and voids. There are both observational and theoretical motives for considering the possibility of their origin tracing back to the events in the very early universe, such as the electroweak phase transition or Inflation. Such a primordial magnetic field (PMF) would remain embedded in the plasma and evolve to persist through the radiation and matter eras, and to the present day. As described in this Chapter, a PMF present in the primordial plasma prior to recombination could help relieve the Hubble tension. A stochastic magnetic field would induce inhomogeneities, pushing the baryons into regions of lower magnetic energy density and speeding up the recombination process. As a consequence, the sound horizon at last scattering would be smaller, which is a necessary ingredient for relieving the Hubble tension. Intriguingly, the strength of the magnetic field required to alleviate the tension is of the right order to also explain the observed magnetic fields in galaxies, clusters of galaxies and voids. These findings motivate further detailed studies of recombination in the presence of PMFs and observational tests of this hypothesis.}

\begin{document} 
\maketitle
\flushbottom


\section{Introduction}

The statistical significance of the Hubble tension, currently just over $5\sigma$, is primarily driven by the difference between the SH0ES measurement of $H_0=73.04 \pm 1.04$ km/s/Mpc using Cepheid calibrated supernova~ \cite{Riess:2021jrx} and the $H_0=67.36 \pm 0.54$ km/s/Mpc obtained from fitting the $\Lambda$CDM model to the Planck CMB data~\cite{Planck:2018vyg}. Other independent measurements tend to re-enforce the tension~\cite{Abdalla:2022yfr}, with a clear trend of all measurements that do not rely on a model of recombination giving a higher $H_0$, in the $69$-$73$ km/s/Mpc range~\cite{Pesce:2020xfe,Wong:2019kwg,Shajib:2019toy,Harvey:2020lwf,Millon:2019slk,Freedman:2020dne}, and estimates that use the standard treatment of recombination giving $H_0$ of around $67-68$ km/s/Mpc~\cite{Ivanov:2019pdj,Aiola:2020azj,Alam:2020sor}. This trend points to a missing ingredient in the standard description of the universe prior and/or during recombination - something that would help reduce the sound horizon at last scattering, $r_\star$. One such ingredient could be a stochastic primordial magnetic field (PMF) embedded in the plasma and generating inhomogeneities in the baryon distribution, or baryon clumping~\cite{Jedamzik:2013gua,Jedamzik:2018itu}. The inhomogeneities make the recombination complete faster~\cite{Peebles:1994xt,Jedamzik:2013gua}, resulting in a smaller $r_\star$.

At the time of writing this Chapter, it remains an open question whether PMFs will play a key role in resolving the Hubble tension. Preliminary investigations~\cite{Jedamzik:2020krr,Thiele:2021okz,Rashkovetskyi:2021,Galli:2021mxk,SPT-3G:2022hvq}, based on a toy-model of baryon clumping~\cite{Jedamzik:2013gua}, have shown that cosmological data is broadly compatible with the presence of PMFs, and they could help relieve the tension to a certain extent. Moreover, the required strength of the PMF appears to be of just the right order of magnitude to help explain all observed galactic, cluster and intergalactic magnetic fields. Needless to say that, if confirmed, this would be a truly exciting development, solving two puzzles at the same time. Achieving a higher level of certainty requires obtaining the exact ionization history from compressible magneto-hydrodynamics (MHD) simulations including the effects of radiation transport that are currently underway~\cite{KJ_TA_2023}. 

In what follows, we start by providing an overview of cosmological magnetic fields, their possible primordial origin and evolution, and their observational signatures. We then describe the physics of recombination in the presence of magnetic fields, introducing the simple model of baryon clumping used in the existing studies. After presenting the current observational status of this model, we discuss the current state of the MHD simulations and prospects for the future.

\section{Primordial magnetic fields and their observational signatures}
\label{sec:pmf}

The PMF hypothesis was first proposed as a possible explanation of the origin of galactic magnetic fields \cite{Hoyle:1958,Zeldovich:1965}. All galaxies, irrespective of their type or age, appear to contain a magnetic field of $\sim$micro-Gauss ($\mu$G) strength that is coherent over the extent of the galaxy. Clusters of galaxies are also known to have magnetic fields of similar strength. The origin of the galactic and cluster fields is not entirely understood \cite{Widrow:2002ud,Widrow:2011hs,Vachaspati:2020blt}. It may well be that initially small fields of strength of $\sim$ $10^{-20}$-$10^{-18}$ G excited by the Biermann battery operating at first structure formation are subsequently amplified by the small-scale and/or large-scale dynamo to reach approximate equipartition of magnetic energy with the turbulent energy in collapsed structures (see~\cite{Brandenburg:2004jv} for a review on astrophysical dynamos). Interestingly, $\mu$G strength fields are observed in high redshift galaxies that would be too young to have gone through the number of revolutions necessary for the large-scale dynamo to work \cite{Athreya:1998}. Still, there is a possibility that the supernova explosions in protogalaxies provided the magnetic seed fields that were later amplified by compression, shearing and stochastic motions \cite{Beck:2013gca,DSeifried:2013zeu}. An alternative explanation is that magnetic fields emerged from the evolution of the early Universe, having been generated either during cosmic phase transitions (e.g. the electroweak transition) \cite{Vachaspati:1991nm} or during inflation \cite{Turner:1987bw,Ratra:1991bn} (see also \cite{Widrow:2002ud,Durrer:2013pga,Subramanian:2015lua,Vachaspati:2020blt}) for reviews).

The pre-recombination universe was fully ionized and, if a PMF was ever generated, dissipation due to magnetic diffusion would be negligible. Dissipation due to the kinetic viscosity damping fluid motions induced by the PMF on small scales 
would, however, significantly drain the energy of the PMF~\cite{Jedamzik:1996wp,Subramanian:1997gi} over
the many e-folds of cosmic expansion between the very early Universe and the recombination era. Notwithstanding this fact, if only the smallest fraction $\sim 10^{-10}$ of a PMF initially in equipartition with the radiation would
survive, it could be sufficient to eliminate the need for dynamo amplification altogether. As discussed below, it could also leave observable signatures in the CMB.

While PMFs have been a subject of continuous study over many decades, there were commonly considered to be unnecessary because less exotic astrophysical explanations could not be ruled out. The interest in PMFs was renewed when evidence emerged of magnetic fields in $\sim$Mpc size voids in the intergalactic space. Observations of TeV blazars by Hess and Fermi~\cite{Neronov:1900zz,Tavecchio:2010mk,Taylor:2011bn} have led to the surprising conclusion that magnetic fields exist in the extra-galactic medium between galaxies with a very large volume filling factor~\cite{Dolag:2010ni}. TeV gamma-rays emitted by the blazar may pair-produce on the diffuse extra-galactic background star light. The resulting electron-positron pairs will subsequently Compton scatter on CMB photons converting them to GeV gamma-rays. Observations of blazars have not found these GeV photons with the flux expected. Though other less understood explanations exist~\cite{Broderick:2011av}, the most straightforward explanation is that the electron-positron pairs were deflected by magnetic fields out of the light cone. A lower limit on the field strength of $B \gtrsim 3\times 10^{-15}$G (assuming a coherence scale of $\sim $kpc) could be inferred. While adding to the case for PMFs, it is not ruled out that such fields could be the result of outflows from galaxies, filling essentially all space with magnetic fields. The synchrotron emission from a few mega-parsec (Mpc) long ridge connecting two merging clusters of galaxies is another observation that is well-matched by simulations based on the PMF hypothesis \cite{Govoni_2019}.

The question of the origin of the magnetic fields in the universe is difficult to settle without the help of further observations. It is hoped that blazar gamma-ray observations by the future Cherenkov Telescope Array mission \cite{CTAConsortium:2018tzg} may raise the lower limit to the $\sim$3-10 pico-gauss (pG) range~\cite{Korochkin:2020pvg}. However, the only way to be certain of the primordial origin of the observed fields is to find their signatures in the CMB. Since the PMF survives Silk damping~\cite{Jedamzik:1996wp,Subramanian:1997gi}, magnetosonic modes on large scales, $\sim$ $1-10$ Mpc, may lead to additional power in the high-$\ell$ tail of the CMB temperature anisotropy spectrum~\cite{Subramanian:1998fn,Subramanian:2002nh,Mack:2001gc,
 Lewis:2004kg,Kahniashvili:2005xe,Chen:2004nf,
 Lewis:2004ef,Tashiro:2005hc,Yamazaki:2006bq,
 Giovannini:2006gz,Kahniashvili:2006hy,Giovannini:2007qn,
 Yamazaki:2010nf,Paoletti:2010rx,
 Shaw:2010ea,Kunze:2010ys,
 Paoletti:2012bb,Ballardini:2014jta,Ade:2015cva,
 Sutton:2017jgr,Zucca:2016iur}. 
Dissipation of magnetic fields before recombination
may lead to spectral distortions~\cite{Jedamzik:1999bm,Zizzo:2005az,Kunze:2013uja,Ganc:2014wia,
Paoletti:2018uic}, and after recombination may lead to an increase of the optical depth~\cite{Sethi:2004pe,Kunze:2013uja,Kunze:2014eka,
Chluba:2015lpa,Ade:2015cva,Paoletti:2018uic}. The PMF may lead to additional polarization anisotropies due to
Faraday rotation~\cite{Durrer:1999bk,Seshadri:2000ky,Mack:2001gc,Subramanian:2003sh,
 Mollerach:2003nq,
 Lewis:2004kg,
 Scoccola:2004ke,
 Kosowsky:2004zh,Kahniashvili:2005xe,Pogosian:2012jd,Kahniashvili:2014dfa,
 Zucca:2016iur,Pogosian:2018vfr} 
The non-Gaussianity of the PMF may lead to a signal in the bi-spectrum or tri-spectrum of the anisotropies~\cite{Brown:2005kr,Seshadri:2009sy,Caprini:2009vk,Cai:2010uw,
 Trivedi:2010gi,Brown:2010jd,Shiraishi:2010yk,Shiraishi:2011dh,Trivedi:2011vt,
 Shiraishi:2013wua,Trivedi:2013wqa,Ade:2015cva}. 
All of the above effects have led to upper limits in the nano-gauss (nG) range. However, as we elaborate in the following subsection, CMB appears to be most sensitive to the generation of small-scale baryonic density fluctuations before recombination, with detectable PMF strength in the $0.01-0.1$ nG range~\cite{Jedamzik:2018itu}. Moreover, as shown in~\cite{Jedamzik:2020krr}, the baryon clumping before recombination may help relieve the Hubble tension.

\section{Recombination with primordial magnetic fields}

A statistically homogeneous and isotropic PMF is characterized by a power spectrum expected to be a smooth function of the Fourier number $k$. On large scales, the universe is a highly conducting plasma with the magnetic field ``frozen-in'' and diluting with the expansion as $B = B_0/a^2$, where $B_0$ is the present day (comoving) strength and $a$ is the scale factor normalized to unity today. On smaller scales, the PMF generates perturbations in the plasma and dissipates its strength. PMFs generated causally in phase transitions generically~\cite{Durrer:2003ja} have a Batchelor spectrum \cite{Batchelor:1959} that monotonically increases with $k$ until a peak at a certain small integral scale \cite{Jedamzik:2010cy}, beyond which it decays. Inflationary magnetogenesis, on the other hand, can produce a range of spectral shapes depending on the particular model (see {\it e.g.} \cite{Durrer:2013pga} for a review), with the simplest models \cite{Turner:1987bw,Ratra:1991bn} predicting a scale-invariant spectrum. CMB constraints are often quoted in terms of the comoving field strength smoothed on the 1 Mpc scale, $B_{\rm 1Mpc}$, which, in many cases, is not a representative measure of the PMF. The more relevant quantity for the baryon clumping effect is the effective comoving strength, $B_{\rm eff}$, quantifying the average magnetic total energy density. For scale-invariant fields, the two measures are essentially the same. For causal fields, however, the difference is quite dramatic, with $B_{\rm eff} \gg B_{\rm 1Mpc}$. Unless specified otherwise, the field strengths quoted in this paper are $B_{\rm eff}$.

Unlike most of the evolution of the magnetized plasma in the early universe, when it is an incompressible fluid,
MHD becomes compressible before recombination for scales well below the photon mean free path $l_{\gamma}\sim 1\,$Mpc\footnote{All cited length scales are comoving.}. Photons are free-steaming over $\sim$ kpc scales and their sole effect is to introduce a drag force on peculiar motions in the CMB rest frame characterized by a linear drag coefficient $\alpha$. The Euler equation for the baryon velocity field takes the form of
\be
{\partial{\bf v} \over \partial t} + ({\bf v}  \cdot {\bf \nabla}){\bf v} + c_s^2 {{\bf \nabla} \rho \over \rho} = - \alpha {\bf v} - {1\over 4 \pi \rho} {\bf B} \times ({\bf \nabla} \times {\bf B})
\label{eq:Euler}
\ee
where the speed of sound, $c_s \approx 2\times 10^{-5}c$, is very small on scales well-below $l_\gamma$, while being much larger, $c_s = \sqrt{1/3}c$, on length scales larger than $l_\gamma$. The first term on the right-hand side (RHS) is the photon drag, while the last term is the Lorentz force due to the PMF,  ${\bf B} \times ({\bf \nabla} \times {\bf B}) = \nabla B^2/2 - ({\bf B} \cdot {\bf \nabla}) {\bf B}$, which pushes baryons into regions of lower magnetic energy density. This force is opposed by the baryon-photon fluid pressure characterized by $c_s^2$. On scales well-below $l_\gamma$, the plasma is free to compress until the baryonic pressure gradients backreact.

Using the continuity equation,
\begin{equation}
{\partial{\rho} \over \partial t} + {\bf\nabla}\bigl(\rho {\bf v}\bigr) = 0 \ ,
\label{eq:continuity}
\end{equation}
one can estimate the amplitude of the generated density fluctuations in a back-of-the-envelope estimate~\cite{Jedamzik:2013gua}. Imagine a stochastic magnetic field at uniform baryon density and negligible velocities initially. These are very realistic initial conditions, since at earlier times the evolving $l_\gamma$ is smaller than the magnetic fluctuations scale $L$, such that the plasma is incompressible, and photon diffusion suppresses any fluid motion. At redshifts $z\sim 10^4-10^6$, depending on the scale, the photon mean free path $l_\gamma$
starts exceeding the scale $L$. At this point Eq.~(\ref{eq:Euler}) applies and there is a significant drop in $c_s$. Initially, only the terms on the RHS of Eq.~(\ref{eq:Euler}) are important and the plasma obtains a terminal velocity
${\rm v}\simeq c_A^2/(\alpha L)$, where $c_A = B/\sqrt{4\pi\rho}$ is the Alfven speed of the plasma. Using Eq.~(\ref{eq:continuity}), it is straigthforward to show that the density fluctuations grow as $\delta\rho /\rho \simeq {\rm v} t/L \simeq c_A^2 t/\alpha L^2$ with time $t$. These density fluctuations become larger with time until either the pressure forces become important in counteracting further compression, or the source magnetic stress term decays. The former happens when the last term on the LHS of Eq.~(\ref{eq:Euler}), $(c_s^2/L) \delta\rho /\rho$, is of the order of the magnetic force term $c_A^2/L$. That is, density fluctuations cannot become larger than $\delta\rho / \rho \lsim (c_A/c_s)^2$. It has been shown by numerical simulation~\cite{Banerjee:2004df} that even in the highly dragged viscous MHD regime magnetic fields decay, albeit slower than during turbulent MHD. Magnetic fields excite fluid motions which in turn get dissipated by the photon drag. It was found that a magnetic structure decays when the Eddy turn-over rate ${\rm v}/L \simeq c_A^2/\alpha L^2$ equals the Hubble rate $H\simeq 1/t$. Entering this into the expression for $\delta\rho /\rho$ one finds that the average density fluctuation cannot exceed unity by much. This leads to the following estimate for the generated density fluctuations:
\begin{equation}
\frac{\delta \rho}{\rho} \simeq {\rm min}\biggl[1,\biggl(\frac{c_A}{c_s}\biggr)^2\biggr] \, .
\end{equation}
Since, at recombination,
\be
c_A = 4.34{\rm km/s}\,\left( B \over 0.03 {\rm nG} \right)
\ee 
and $c_s = 6.33 {\rm km/s}$, even fairly weak fields may generate order-unity density fluctuations on small scales. This simple estimate has subsequently been confirmed by full numerical simulations~\cite{Jedamzik:2018itu}. 

As the Universe expands, density fluctuations on ever larger scales are generated and subsequently decay. Causally generated PMFs loose a significant fraction of their power with each e-fold as the peak of the Batchelor spectrum keeps moving to larger scales and weaker fields according to $B(L) = B_0 (L/L_0)^{-5/2}$. Scale-invariant fields, on the other hand, are largely unaffected by the evolution. As discussed above, that magnetic fields and the associated density fluctuations decay when the Eddy turnover time equals the Hubble time. This can be used to derive a correlation between the magnetic field strength and the scale of the smallest magnetic structure that survives until just before recombination~\cite{Banerjee:2004df}:
\begin{equation}
B_{\rm rec}\lsim 80\, p{\rm G} \biggl(\frac{L}{\rm kpc}\biggr)\, ,
\label{eq:correlation}
\end{equation}
where approximate equality is attained for fields which do undergo dissipation in the first place. For fields which are so weak and/or are on such large scales that the Eddy turnover time never reaches the Hubble time, no dissipation ever occurs and no correlation can be derived (hence the inequality sign). Thus, quite generally, for sufficiently strong fields, $B\sim 10-100$ pG, inhomogeneities are generated on $\sim 0.1-1$ kpc scales shortly before recombination. Further dissipation occurs across recombination as a result of the sudden drop in the free electron density. This leads to a factor of $6.2$ drop in the strength of the causal PMF, while, again, a scale-invariant field strength is not affected. After recombination, the PMF evolution is almost entirely halted or at least significantly slowed down as recent numerical simulations show~\cite{KJ_TA_2023}. 
The post-recombination field strength is the one relevant for the subsequent structure formation and is commonly referred to as ``pre-collapse''. For causal fields, the resultant pre-collapse correlation is given by $B_0\simeq 5\,p{\rm G} (L/{\rm kpc})$~\footnote{It is noted that there is no conflict between the statement that the magnitude of $B_{\rm eff}$ diminishes by a factor $6.2$ during recombination, on the one hand, and Eq.~\ref{eq:correlation}
and this relation, on the other hand, as the coherence length $L$ is increasing during the dissipation.}. 
As the differences between various measures of the PMF are model-dependent and can be a source of confusion, we provide a glossary in Table~\ref{tab:B}.  It has been shown by detailed simulations~\cite{Dolag:99,Dolag:02} that, irrespective of their coherence length, pre-collapse magnetic fields of $\sim$5 pG lead to final post-collapse cluster fields of $\sim$ $\mu$G with Faraday rotation measures in good agreement with observations.

\begin{table}[tbp]
\centering
\begin{tabular}{c|c|c|c|c}
Spectrum & $B_{\rm eff}$ at $z_\star$ &  $B_{\rm 1Mpc}$ at  $z_\star$ &  $B_{\rm eff}$ at $z=10^6$ & $B_{\rm eff}^{\rm pre-collapse}$ \\
\hline
Scale-invariant & 0.05 & 0.05 & 0.05  & 0.05 \\
Batchelor & 0.05 & $10^{-9}$ & 5  &  0.008  \\
 \hline
\end{tabular}
\caption{\label{tab:B} Different measures (in nG) of a PMF that has an effective comoving magnetic strength of $B_{\rm eff} = 0.05$ nG at the redshift of last-scattering $z_\star$. All the measures are the same for scale-invariant fields, but differ significantly for a causally generated PMF with a Batchelor spectrum. 
} 
\end{table}

The magnetically induced inhomogeneities are on scales much too small (i.e. $\ell \sim 10^6-10^7$) to be observed directly as a contribution to CMB anisotropies. However, the baryon clumping has a prominent effect on the rate of recombination, speeding it up and resulting in a smaller sound horizon at photon-baryon decoupling, $r_\star$. The speed up of recombination follows from the recombination rate being proportional to the square of the electron density $n_e$. Since, generally, $\langle n_e^2\rangle > \langle n_e\rangle^2$ in a clumpy universe, where $\langle \rangle$ denotes spatial average, the recombination rate is enhanced compared to $\Lambda$CDM.  A smaller $r_\star$ raises the value of $H_0$ inferred from the very accurately measured angular size of the sound horizon, helping reduce the Hubble tension.

\section{Inhomogeneous recombination}
\label{sec:pdf}

The evolution of the cosmic electron density depends on the full details of the probability distribution function (PDF) $p(\Delta)$ to find baryons at density $\Delta \langle\rho \rangle$, where $\langle\rho \rangle$ is the average baryonic density and $\Delta$ is an enhancement factor. The shape of the baryon
density PDF is currently unknown, and neither is its evolution before recombination, which could be substantial, particularly for blue Batchelor spectra. In the absence of knowledge of the PDF,~\cite{Jedamzik:2018itu} and~\cite{Jedamzik:2020krr} utilized a simple three-zone model (see Sec.~\ref{sec:3zone}) to describe the distribution of baryons. In the three-zone model, the amplitude of baryon inhomogeneity is controlled by the clumping factor
\begin{equation}
b \equiv \biggl( \frac{\langle\rho^2 \rangle-\langle\rho \rangle^2}{\langle\rho \rangle^2} \biggr)
\end{equation}
which sets the second moment of the PDF, while the higher order moments are effectively determined by the choice of the three-zone model parameters. As we show next, the higher moments, and the third moment in particular, play a very important role in determining the enhancement of the recombination rate.

The evolution of the electron number density around recombination is given by
\begin{equation}
\frac{{\rm d}n_e}{{\rm d}t} + 3 H n_e = -C\bigl(\alpha_e n_e^2-\beta 
n_{H^0} e^{-\frac{h\nu_{\alpha}}{T}}\bigr) \ ,
\label{ne_rate}
\end{equation}
where $n_e$ is the free electron density, $n_{H}$ is the total matter number density (protons and hydrogen), and $n_{H^0} = n_{H} - n_e$ is the density of neutral hydrogen. For simplicity, to illustrate the importance of the higher moment, we consider a universe with only hydrogen ({\it i.e.} no helium). Since Eq.~(\ref{ne_rate}) is quadratic in $n_e$, and since $\langle n_e^2 \rangle > \langle n_e\rangle^2$ in inhomogeneous universes, we expect the recombination rate to be enhanced when inhomogeneities exist. It would be tempting to solve the above equation by the introduction of a clumping factor given by the underlying density distribution, {\it i.e.} $b = (\langle n_H^2\rangle - \langle n_H\rangle^2)/\langle n_H\rangle^2$, such that the recombination term would be replaced by $\alpha_e (1+b)\langle n_e\rangle^2$ and the effect of inhomogeneities would be entirely described by the first two moments of the distribution, {\it i.e.} the average density $\langle n_H\rangle$ and the variance $\langle n_H^2\rangle$. However, such a procedure would be incorrect as it would assume that $\langle n_e^2\rangle\sim \langle n_H^2\rangle$. This is not the case if different density regions have different ionization fractions $\chi_e = n_e/n_H$. Rather, as shown below, the average ionization depends on all the moments of the underlying probability distribution of the density fluctuations $P(n_H)$.

Consider detailed balance, {\it i.e.} set the RHS of Eq.~(\ref{ne_rate}) to be approximately zero. This assumption holds fairly during the middle of recombination because both the recombination and the photoionisation rates are much larger than the Hubble rate, but is increasingly inaccurate towards the end of recombination. The equilibrium ionization fraction can then be derived as
\begin{equation}
\chi_e^2 = \frac{\beta_e}{\alpha_e}\frac{1}{n_H^2}(n_H-n_e)e^{-\frac{h\nu_{\alpha}}{T}} =  \frac{\beta_e}{\alpha_e}\frac{1}{n_H}e^{-\frac{h\nu_{\alpha}}{T}}(1-\chi_e) .
\end{equation}
Let us now allow for inhomogeneities in the total baryon density by defining
\begin{equation}
n_H = \Delta \langle n_H\rangle ,
\end{equation}
where $\langle...\rangle$ denotes the ensemble average, which we will assume to be the same as the average over any sufficiently large volume. Then we can write
\begin{equation}
\chi_e^2 =  \frac{\beta_e}{\alpha_e}\frac{1}{\Delta \langle n_H\rangle}e^{-\frac{h\nu_{\alpha}}{T}}(1-\chi_e) = {A \over \Delta} (1-\chi_e)  ,
\label{chi}
\end{equation}
where we have defined 
\begin{equation}
A \equiv \frac{\beta_e}{\alpha_e}\frac{1}{\langle n_H\rangle}e^{-\frac{h\nu_{\alpha}}{T}} .
\end{equation}
From (\ref{chi}), we can write the local ionized fraction as
\begin{equation}
\chi_e (\Delta)= \frac{A}{2\Delta}\biggl(1+\frac{4\Delta}{A}\biggr)^{1/2}-\frac{A}{2\Delta} .
\end{equation}
In the homogeneous case, $\Delta=1$, we have
\begin{equation}
\chi_e^0 (\Delta=1)= \frac{A}{2}\biggl(1+\frac{4}{A}\biggr)^{1/2}-\frac{A}{2} .
\end{equation}
Note that in the limit $A \gg 1$ we have a full ionisation, i.e. $\chi_e \rightarrow 1$, while in the opposite limit, $A \ll 1$, we have $\chi_e \rightarrow 0$.

For consistency, the inhomogeneity parameter $\Delta(x)$ (x are space coordinates) must fulfill
\begin{equation}
\frac{1}{V_{tot}}\int {\rm d}V\Delta (x) = 1
\label{Delta}
\end{equation}  
where the integration is over the total spatial volume $V_{tot}$. For simplicity, as assumed earlier, we will take averages over this volume to be the same as ensemble averages and the above constraint can be simply written as $\langle\Delta\rangle=1$. The average electron density is given by
\begin{equation}
\langle n_e\rangle = \langle \chi_e(\Delta)\Delta \langle n_H\rangle\rangle \ ,
\end{equation}
and the average ionzation fraction is
\begin{equation}
\langle\chi_e\rangle \equiv \frac{\langle n_e\rangle}{\langle n_H\rangle} =  \frac{A}{2} \langle \sqrt{1+\frac{4\Delta}{A}}-1 \rangle .
\label{ave-chi}
\end{equation}
Note that this is not the same as the average of Eq. (\ref{chi}), where $\chi_e = n_e/n_H$. Namely, $\langle n_e\rangle/\langle n_H\rangle \neq \langle n_e/n_H\rangle$. The relevant quantity for us is actually $\langle n_e\rangle$, which is what appears in the CMB calculations, hence the relevant quantity is the one in Eq.~(\ref{ave-chi}).

One can now show that, for arbitrary distribution functions $P(n_H)$ or $\tilde{P}(\Delta )$ fulfilling the
integral constraint
\begin{equation}
\langle\Delta\rangle \equiv \int {\rm d}\Delta \Delta \tilde{P}(\Delta ) = 1 ,
\end{equation}
the $\langle\chi_e\rangle$ of Eq.~(\ref{ave-chi}) is smaller than $\chi_e^0$. Namely, that
\begin{equation}
\big\langle \sqrt{1+\frac{4\Delta}{A}} \big\rangle < \sqrt{1+\frac{4}{A}} \ .
\end{equation}
This follows from Jensen's inequality~\cite{10.1007/BF02418571} which states that the expectation value $\langle g(X)\rangle$ of any concave function $g(X)$ of a random variable $X$ is smaller than $g(\langle X\rangle)$. In our case, the function $\sqrt{1+4\Delta/A}$ is certainly concave and hence we have our result.

For smaller density perturbations one can also see this without using Jensen's theorem by introducing $\delta \equiv \Delta-1$, such that $\langle\delta\rangle=0$, and performing a Taylor expansion under the assumption that $\delta < 1$. Then,
\begin{equation}
\big\langle \sqrt{1+\frac{4}{A}+\frac{4\delta}{A}} \big\rangle \approx \sqrt{1+\frac{4}{A}} - \gamma_2 \langle \delta^2 \rangle + \gamma_3 \langle \delta^3 \rangle - ... ,
\end{equation}
where all $\gamma_i > 0$. One can see that the leading second order term is negative (the function is concave), as well as all even order moment terms, while the odd moments come with a positive signs. This means, {\it e.g.}, that PDFs with a large third moment could reduce the impact on the ionized fraction which is a trend we have observed while experimenting numerically with different three-zone models. Ultimately, the full redshift evolution of the baryon PDF and the associated ionized fraction will be determined exactly from MHD simulations that are currently in progress.

\subsection{The three-zone model}
\label{sec:3zone}

In the absence of the baryon density PDF derived from MHD simulations, when discussing the current observational constraints on the PMF-enhanced recombination, we will use the simple three-zone model introduced in~\cite{Jedamzik:2013gua}.  The model is described by the density parameters $\Delta_i$ and volume fractions $f_V^i$ in each zone. Baryon densities in the individual zones are simply given by $n_b^i = \langle n_b\rangle \Delta_i$. Parameters $\Delta_i$ and $f_V^i$ have to fulfil the following constraints:
\begin{eqnarray}
\sum_{i=1}^3 f_V^i =1, \ \sum_{i=1}^3 f_V^i\Delta_i = 1, \ 
\sum_{i=1}^3 f_V^i\Delta_i^2 = 1 + b\, ,
\end{eqnarray}
{\it i.e.} the total volume fraction is one and the three-zone model has average density $\langle n_b\rangle$ and clumping factor $b$.
This leads to three constraints for six free parameters ${f_V^i,\Delta_i}$, such that one may choose three parameters freely. To obtain the average ionization fraction $\langle \chi_e \rangle$, one computes the ionization fraction in each of the zones and take the average, {\it i.e.}
\begin{equation}
\langle \chi_e \rangle = \sum_{i=1}^3 f_V^i\Delta_i \chi_e^i \ .
\end{equation}

In~\cite{Jedamzik:2013gua}, the parameters were chosen to be $f_V^2 = 1/3$, $\Delta_1 = 0.1$ and $\Delta_2 = 1$, which we will subsequently refer to as the M1 model. In~\cite{Jedamzik:2020krr}, for comparison, we introduced a second model, M2, that had $f_V^2 = 1/3$, $\Delta_1 = 0.3$ and $\Delta_2 = 1$. It is not clear at this point if either of these models provides a good representation of the actual baryon PDF, but we note that M2 has a larger third moment and, for reasons discussed earlier, results in a lesser reduction of the average ionized fraction for the same value of the clumping factor $b$.

\section{The impact of baryon clumping on CMB spectra}
\label{sec:cmb}


The discussion in this and the subsequent Sections is largely based on the results obtained in~\cite{Galli:2021mxk}, where the impact of clumping on the CMB spectra and, in particular, their effect on CMB polarization were studied. As previously discussed, inhomogeneous recombination completes sooner, which shifts the peak of the visibility function to an earlier epoch. This lowers $r_\star$, defined as the comoving sound horizon at the peak of the visibility function, which has the effect of shifting the acoustic peaks in CMB spectra to smaller angular scales. One can compensate for the shift with a larger value of $H_0$, which may help in relieving the Hubble tension.

In addition to the shift of the peaks, earlier recombination means that CMB polarization is produced at an earlier epoch and, as a consequence, has a larger overall amplitude. This is because the value of the speed of sound $c_s$ (on scales where the baryon-photon fluid is tightly coupled) was larger at earlier times, and the amplitude of polarization is set by the quadrupole of temperature anisotropy, which is derived from the dipole, which is set by the time derivative of the monopole, which in turn is proportional to $c_s$~\cite{Zaldarriaga:1995gi}. 

\begin{figure*}[tbph!]
  \centering
\includegraphics[width=0.85\textwidth]{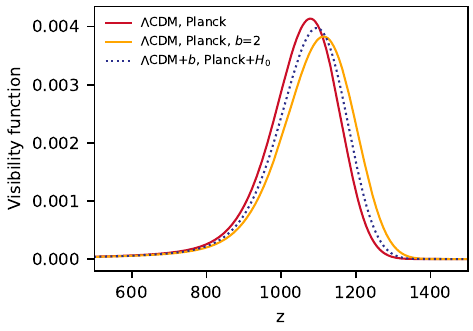}
\caption{Impact of baryon clumping on the CMB visibility function. We compare the visibility functions in the Planck best-fit \LCDM\ model (solid red line), an M1 model with $b=2$ and all cosmological parameters set to the best-fit \lcdm\ (solid orange line), and the \lcdmb Planck+SH0ES best-fit M1 model (dotted blue line). One can see that clumping shifts the peak of the visibility function to earlier times, and increases its width. Figure reproduced from~\cite{Galli:2021mxk}.}
\label{fig:visibility}
\end{figure*}

Clumping also broadens the visibility function, which is a consequence of the overdense baryon pockets recombining earlier and underdense baryon pockets recombining later. This broadening tends to further enhance polarization, because of the period of time during which polarization can be generated is longer. 

The two effects, the shift of the peak and the broadening of the visibility function, are illustrated in Fig.~\ref{fig:visibility}, which compares the visibility functions in the \lcdm\ model, with an M1 model with $b=2$ with all other parameters kept the same, and with the M1 model (\lcdmb) that best fits the Planck CMB data combined with the SH0ES prior on $H_0$. The broadening effect is apparent from the lower peak, since the visibility function is normalized to integrate to unity. We note that, while the general trends in the visibility function are common to all clumping models, the quantitative details are dependent on the shape and the evolution of the baryon density PDF. 

\begin{figure*}[tbph!]
  \centering
\includegraphics[angle=0,width=0.95\textwidth]{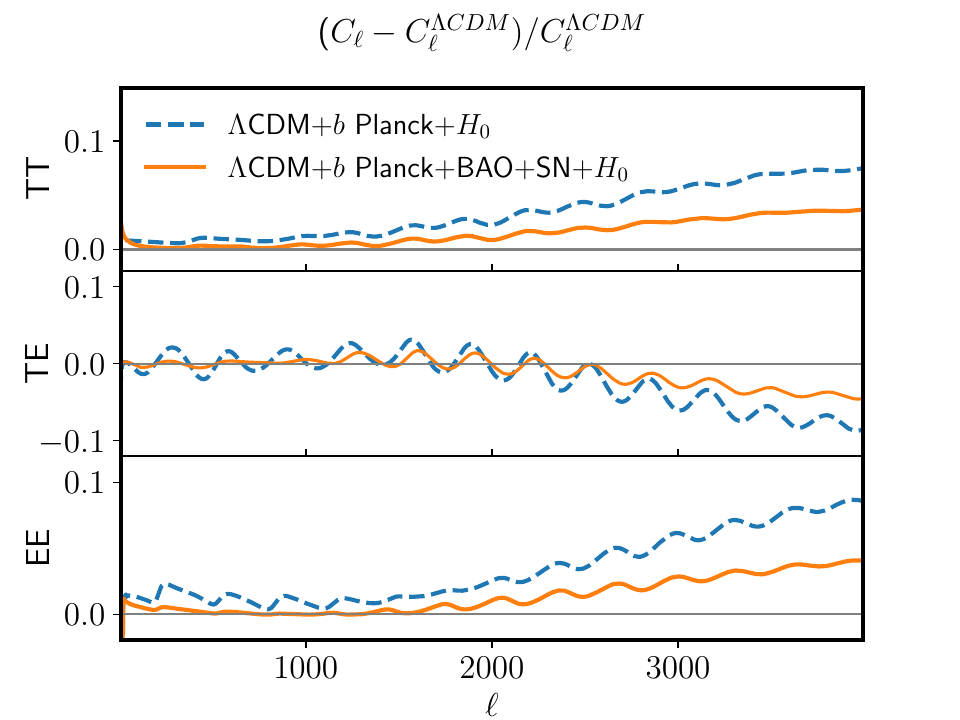}
\caption{Impact of baryon clumping on CMB power spectra. Shown is the relative difference between the CMB spectra in the Planck best-fit $\Lambda$CDM model and the M1 models best-fit to Planck+SH0ES (blue dashed line) and Planck+BAO+SN+SH0ES (orange solid line). In the case of TE, which is positively or negatively correlated depending on $\ell$, we avoid divisions by zero by computing $(C_\ell - C_\ell^{\Lambda CDM})/C_\ell^{ref}$, where $C_\ell^{ref}$ is the absolute value of $C_\ell^{\Lambda CDM}$ convolved with a Gaussian of width $\sigma_\ell=100$ centred at $\ell$. The plot illustrates the importance of small-scale CMB anisotropies for constraining baryon clumping. Figure reproduced from~\cite{Galli:2021mxk}.}
\label{fig:pmf_cmb_spectra}
\end{figure*}

Another significant effect of clumping on CMB spectra comes from a modification of the Silk damping scale $r_D$. Here, there are three competing effects. Firstly, $r_D$ decreases due to an overall earlier completion of recombination. This decrease, however, could be negated by the fact that an earlier broad helium recombination in clumping models results in a smaller electron density available at the later stages of recombination, and due to the broadening of the visibility function. Since much of the Silk damping occurs right at recombination, where the visibility function is of order unity, the details of the visibility function play an important role. The first effect, due the overall shift to higher redshifts, would reduce the Silk damping effect on CMB spectra, as it pushes the onset of the damping tail to higher multipoles $\ell$. The second and third effects, however, can also be important, and the balance between them is model-dependent and varies with the clumping factor. In the best-fit M1 model, we find that there is less Silk damping compared to \lcdm. But, in the best-fit M2 model (see Sec.~\ref{sec:3zone}), the Silk damping is virtually identical to that in the best-fit \lcdm. Also, in M1, at (observationally disallowed) high values of $b$, the Silk damping is actually enhanced. Additional discussion of the evolution of the damping scale as a function of $b$ in a different clumping models can be found in~\cite{Rashkovetskyi:2021}. The evolution of magnetically induced clumping is unknown at present and is a significant source of uncertainty in observational constraints on the PMF. For example, if clumping was stronger at $z\sim 3000$ than at $z\sim 1200$, the helium recombination could be the dominant effect, inducing more Silk damping.

The reduction in Silk damping present in the M1 model is illustrated in Fig.~\ref{fig:pmf_cmb_spectra}, which compares the CMB spectra in the Planck best-fit $\Lambda$CDM to those in the Planck+$H_0$ and Planck+BAO+SN+$H_0$ best-fit M1 models, where $H_0$ denotes the measurement by SH0ES. The enhancements in the temperature (\TT) and E-mode polarization (\EE) spectra at high $\ell$ is due to the smaller $r_D$. The same enhancement is also seen in the temperature-polarization cross-correlation (\TE), which is negative due to the anticorrelation between T and E at high $\ell$. This illustrates the fact that high resolution CMB measurements will be a key discriminant in constraining inhomogeneous recombination models\footnote{Fig.~\ref{fig:pmf_cmb_spectra} also shows that at $\ell \lesssim 20$ the polarization is reduced, which is due to the lower best fit value of the optical depth $\tau$.}. Also, as first pointed out in~\cite{Galli:2021mxk}, one needs the full combination of \TT, \TE\ and \EE\ CMB spectra, as neither \TT\ nor \TE+\EE\ on their own are able to break degeneracies between primordial power spectrum index $n_s$, the amplitude $A_s e^{-2\tau}$, and other cosmological parameters, required for placing tight constraints on $b$. 

\section{Relieving the Hubble tension with PMFs}
\label{sec:constraints}

The extent to which PMFs can help relieve the Hubble tension is limited by the fact that reducing the sound horizon $r_\star$, by itself, can at best raise the value of $H_0$ to $\sim 70$ km/s/Mpc without causing new tensions between CMB, BAO and the large-scale structure clustering. In what follows, we first reproduce the general argument illustrating this point, originally made in \cite{Jedamzik:2020zmd}, before describing the current observational status of the M1 baryon clumping model.

\subsection{Why reducing $r_\star$ helps, but can not fully resolve the Hubble tension by itself}
\label{sec:bao}

\begin{figure*}[tbph!]
  \centering
    \includegraphics[width=0.9\textwidth]{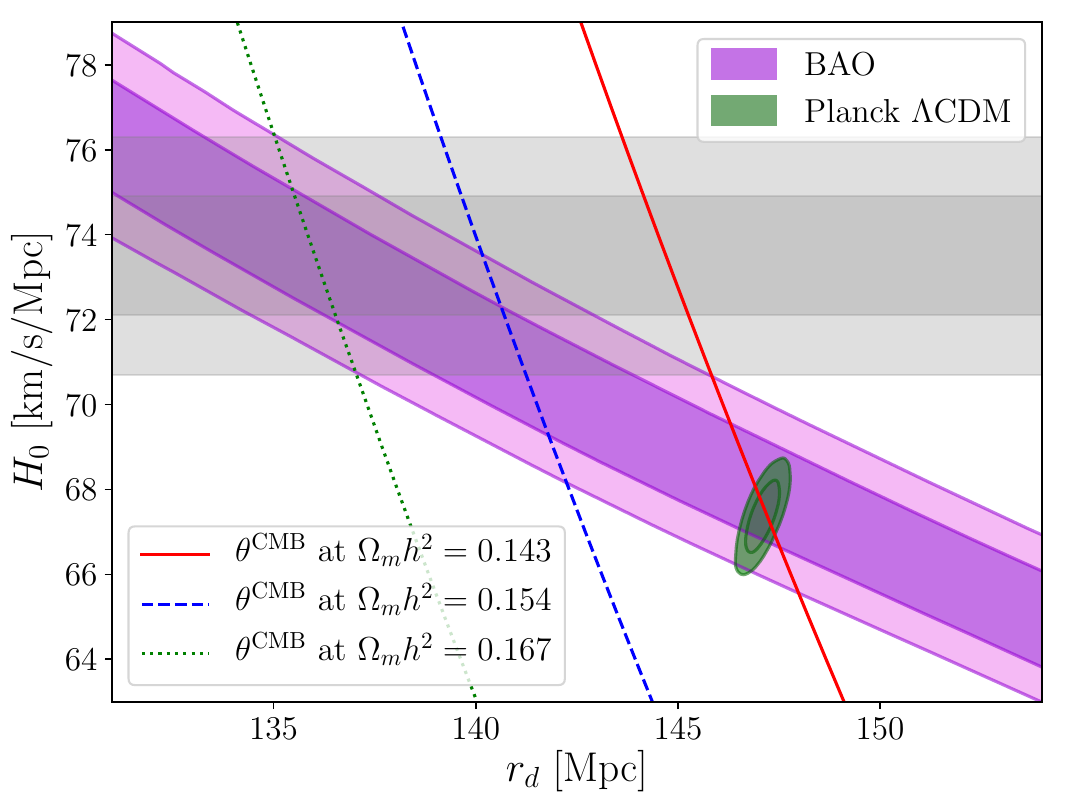}
 \caption{  \label{fig:cmb-bao2}
A plot illustrating that achieving a full agreement between CMB, BAO and SH0ES through a reduction of $\rd$ requires a higher value of $\omm$. Shown are the CMB lines of degeneracy between the sound horizon $\rd$ and the Hubble constant $H_0$ at three different values of $\omm$: $0.143$, $0.155$ and $0.167$. Also shown are the $68\%$ and $95\%$ CL bands derived from the combination of all current BAO data marginalized over $\omm$, the $\Lambda$CDM based bounds from Planck and the determination of the Hubble constant by SH0ES.
}
\end{figure*}

\begin{figure*}[htbp] 
  \centering
\includegraphics[scale=0.8]{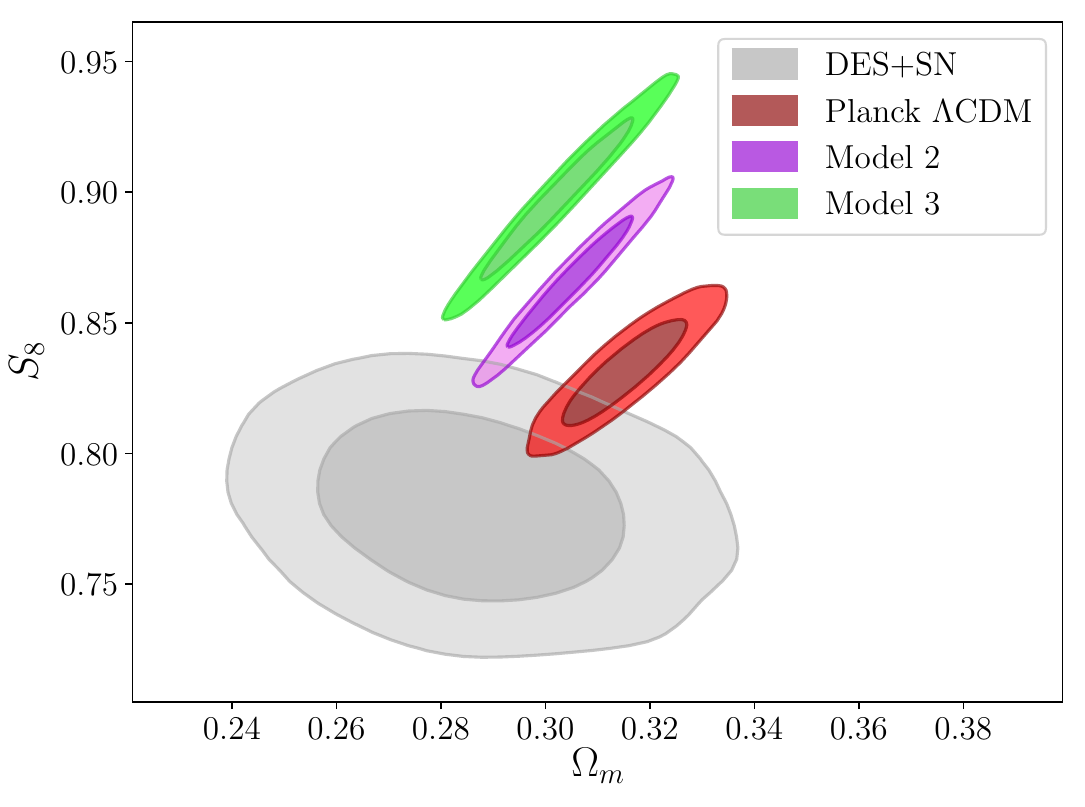}
\caption{
The $68\%$ and $95\%$ confidence level bounds on $S_8$ and $\Om$. Shown are the constraints derived by fitting the $\Lambda$CDM model to a joint dataset of DES and SN and to Planck, along with the contours for Model 2 and Model 3 corresponding to the $\omm = 0.155$ and  $\omm=0.167$ $\rd-H_0$ lines in Fig~\ref{fig:cmb-bao2}. Figure reproduced from \cite{Jedamzik:2020zmd}.
} 
\label{fig:S8-Om}
\end{figure*}

The acoustic peaks in CMB anisotropy spectra provide a very accurate measurement of the angular size of the sound horizon at recombination,
\begin{equation}
\theta_\star = \frac{r_{\star}}{D(z_{\star})} 
\, ,
\label{eq:theta*}
\end{equation}
where $D(z_{\star})$ is the comoving distance from a present day observer to the last scattering surface, where $z_{\star}$ is the redshift of the peak of the visibility function. In a given model, $r_\star$ and $D(z_{\star})$ can be determined from $r_{\star} = \int_{z_{\star}}^\infty c_s(z) {\rm d} z /H(z)$ and $D(z_{\star})=\int^{z_\star}_0 c \ {\rm d} z / H(z)$, where $c_s(z)$ is the sound speed of the photon-baryon fluid, $H(z)$ is the redshift-dependent cosmological expansion rate and $c$ is the speed of light. To complete the prescription, one also needs to determine $z_\star$ using a model of recombination.

The sound waves responsible for the acoustic peaks in the CMB spectra are also imprinted in the galaxy power spectra as Baryon Acoustic Oscillations (BAO) with a minor difference in this standard ruler. Rather than $r_\star$, which is the sound horizon at photon decoupling, the scale imprinted in the BAO is the sound horizon at the baryon decoupling $\rd$, also known as the ``cosmic drag'' epoch when the photon drag on baryons becomes unimportant. The latter takes place at a slightly lower redshift than recombination, so that $r_{\rm d} \approx 1.02 r_{\star}$ in \lcdm. The proportionality factor between the two sound horizons does not change appreciably in alternative recombination scenarios. 

While the difference between $r_\star$ and $\rd$ is small, the difference in redshifts at which the acoustic features in the CMB and BAO are observed is significant, with the latter measured at $0 \lesssim z \lesssim 2.5$ accessible to galaxy redshift surveys. The angular scale of the BAO feature measured using galaxy correlations in the transverse direction to the line of sight is
\begin{equation}
\theta_\perp^{\rm BAO} (z_{\rm obs}) \equiv \frac{r_{\rm d}}{D(z_{\rm obs})} \, ,
\label{eq:thetaBAO}
\end{equation}
where $z_{\rm obs}$ is the redshift of the correlated galaxies.

Let us now consider Eqs.~(\ref{eq:theta*}) and (\ref{eq:thetaBAO}) while remaining agnostic about the particular model that determines the sound horizon. Namely, let us treat $r_\star$ as an independent parameter and assume $r_{\rm d} = 1.02 r_{\star}$. Let us also assume that after recombination the expansion of the universe is well-described by the \lcdm\ model and, for simplicity, ignore the contribution of radiation to the  distance integrals $D(z_\star)$ and $D(z_{\rm obs})$. Then, for the CMB acoustic feature, we can write
\begin{equation}
\theta_{\star} = {r_\star \over 2998\, {\rm Mpc}} \left( \int_0^{z_{\star}}
\frac{{\rm d}z}{\omega_m^{1/2}\sqrt{(1+z)^3 + h^2/\omega_m -1}} \right)^{-1},
\label{eq:rstar}
\end{equation}
where $\omega_{\rm m} = \om$ is the matter density today, $\Omega_m$ is the fractional matter density, $h$ is $H_0$ in units of 100 km/s/Mpc, and $2998\, {\rm Mpc} = c/100{\rm km/s/Mpc}$. An analogous equation for the BAO is obtained by replacing $(r_{\star}, \theta_\star ,z_{\star})$ with $(r_{\rm d},\theta_\perp^{\rm BAO},z_{\rm obs})$. 

For a given $\omega_{\rm m}$, and with the precisely measured $\theta_{\star}$, Eq.~(\ref{eq:rstar}) defines a line in the $\rd$-$H_0$ plane\footnote{From here on we will use $\rd$ to represent the acoustic feature in both the CMB and BAO}. Similarly, a BAO measurement at each different redshift also defines a respective line in the $\rd$-$H_0$ plane. However, the slopes of the CMB and BAO lines are very different due to the vast difference in redshifts at which the standard ruler is observed, $z_{\star}\approx 1100$ for CMB {\it vs} $z_{\rm obs} \sim 1$ for BAO. This is illustrated in Fig.~\ref{fig:cmb-bao2} that shows the $\rd-H_0$ degeneracy line for CMB at three different values of $\omega_{\rm m}$, as well as the $68\%$ and $95\%$ confidence level (CL) regions derived from the combination of all presently available BAO observations while treating $\rd$ as as an independent parameter and marginalizing over $\omega_{\rm m}$ (see \cite{Pogosian:2020ded} for details). Also shown are the \lcdm\ constraint from Planck (in red) and the SH0ES determination of $H_0$ (the grey band).  On can see from Fig.~\ref{fig:cmb-bao2} that BAO is consistent with either Planck or SH0ES, depending on the value of $\rd$, but the latter two are in clear disagreement with each other for the Planck-\lcdm-preferred $\omega_{\rm m}\approx 0.143$. 

One can reconcile Planck with SH0ES by reducing $\rd$ and ``moving up'' the CMB $\rd-H_0$ degeneracy line. However, doing so at a fixed $\omega_{\rm m}$ would quickly move the values of $\rd$ and $H_0$ out of the purple band in Fig.~\ref{fig:cmb-bao2} creating a tension with BAO. In order to reconcile CMB, BAO and SH0ES at the same time, one needs to reduce $\rd$ and increase the matter density, as illustrated in the Figure. In particular, a full resolution of the tension would require $\omega_{\rm m} \approx 0.167$, which is substantially higher than the best-fit \lcdm\ value. A larger matter density results in more clustering quantified by the $S_8$ parameter. It is defined as $S_8 \equiv \sigma_8(\Omega_m/0.3)^{0.5}$, where $\sigma_8$ is the matter clustering amplitude on the scale of $8 \ h^{-1}$Mpc. A larger $S_8$ would exacerbate the already existing mild tension between its Planck best-fit value and that measured from weak gravitational lensing of galaxies~\cite{Kilo-DegreeSurvey:2023gfr}.

The reason why simply reducing $r_\star$ would not fully solve the Hubble tension is seen in Fig.~\ref{fig:S8-Om}. The figure shows the  $68\%$ and $95\%$ CL constraints on $S_8$ and $\Om$ by DES supplemented by the Pantheon SN sample \cite{Scolnic:2017caz}. Also shown is the Planck \lcdm\ best fit, corresponding to $\omega_{\rm m}  = 0.143$, as well as the two cases corresponding to $\omega_{\rm m}  = 0.155$ (Model 2) and $\omega_{\rm m} =0.167$ (Model 3). The constraints for the latter two were obtained by simultaneously fitting BAO and CMB acoustic peaks at the corresponding fixed values of $\omega_{\rm m}$. One can see that when attempting to bring CMB, BAO and SH0ES in agreement with each other by reducing $\rd$, one increases the $S_8$ tension.

Thus, any proposed solution to the Hubble tension that amounts to reducing $\rd$ without other significant changes to \lcdm\ would not be able to raise the CMB-extracted value of $H_0$ above $70$ km/s/Mpc. Baryon clumping belongs to that category of models, hence we do not expect it to fully resolve the Hubble tension. But it could still help relieve it to a statistically acceptable level.

\subsection{Observational status of the M1 baryon clumping model}
\label{sec:m1_constraints}

\begin{figure*}[tbph!]
  \centering
    \includegraphics[width=0.8\textwidth]{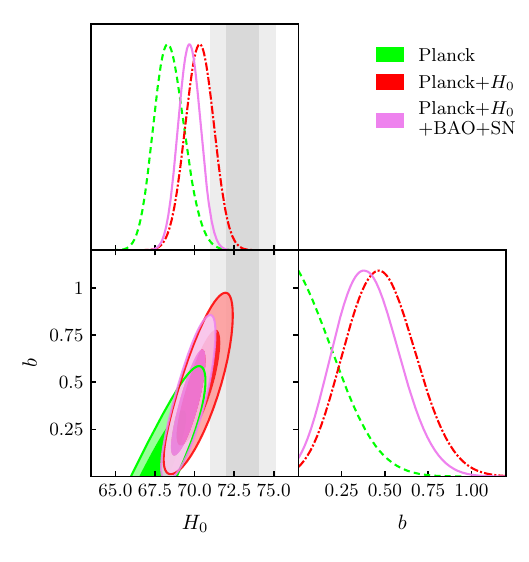}
  \caption{  \label{fig:update1}
Constraints on the clumping factor $b$ and $H_0$ derived from \planck\ (lime), the combination of \planck\ with the SH0ES prior on $H_0$ (red), and the combination of the \planck, SH0ES, BAO and SN data (violet). With the $H_0$ prior, there is a preference for a nonzero clumping at $\sim$3$\sigma$ level.
}
\end{figure*}

\begin{figure*}[tbph!]
  \centering
\includegraphics[width=0.9\textwidth]{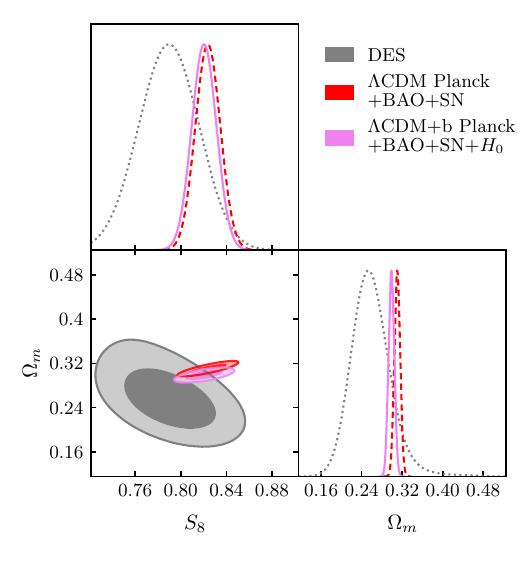}
\caption{\label{fig:s8}Joint constraints in the $S_8$-$\Omega_m$ plane from Planck+BAO+SN in \lcdm\ (red), and in the M1 model fit to Planck+BAO+SN with the SH0ES-based $H_0$ prior (violate). The \lcdm\ based  constraints from the DES Y1 data \cite{Abbott:2017wau} are shown in grey. There is a slight relief of the tension in M1 due to $\Omega_mh^2$ remaining largely the same as in the best-fit \lcdm{} model, while $h$ is increased.
}
\end{figure*}

In the absence of the detailed ionized fraction evolution obtained from MHD simulations with a PMF (see Sec.~\ref{sec:mhd}), the M1 three-zone model~\cite{Jedamzik:2013gua} was used in \cite{Jedamzik:2020zmd,Thiele:2021okz,Rashkovetskyi:2021,Galli:2021mxk,SPT-3G:2022hvq} for exploring the ability of baryon clumping to resolve the Hubble tension. Below we summarize the current observational constraints on the M1 model.

Fig.~\ref{fig:update1} shows the marginalized posteriors for the clumping factor $b$ and $H_0$ from Planck, Planck+SH0ES and Planck+SH0ES+BAO+SN, where we used the eBOSS DR16 BAO compilation from~\cite{Alam:2020sor} and SN stands for the Pantheon supernovae sample~\cite{Scolnic:2017caz}. We see that Planck by itself shows no preference for clumping, with $b < 0.47$ at 95\% CL and only a small increase in the best-fit $H_0$. Adding the SH0ES prior (H0) to Planck gives $b=0.48 \pm 0.19$ and $H_0 = 70.32 \pm 0.85$ km/s/Mpc. Adding the BAO and SN data results in $b=0.40^{+0.15}_{-0.19}$ and $H_0 = 69.68 \pm 0.67$ km/s/Mpc, a reduction due to the general difficulty in maintaining the agreement between CMB and BAO while decreasing the sound horizon, as discussed in the previous subsection.

Additional constraints on baryon clumping can be derived by adding the high-resolution CMB temperature and polarization data from the Atacama Cosmology Telescope fourth data release (\actdr) \cite{Choi:2020ccd} and the South Pole Telescope Third Generation (SPT-3G) 2018 data \cite{Dutcher:2021vtw,SPT-3G:2022hvq}. As discussed in Sec.~\ref{sec:cmb} and demonstrated in Fig.~\ref{fig:pmf_cmb_spectra}, baryon clumping impacts the Silk damping tail of the CMB spectrum, making the spectra at multipoles $\ell > 2000$ a powerful probe of baryon clumping. Combining Planck and ACT yields $b < 0.34$ at 95\% CL~\cite{Thiele:2021okz,Galli:2021mxk}, while Planck+SPT gives $b<0.38$ at 95\% CL~\cite{SPT-3G:2022hvq} in the M1 model.

Many models that resolve the Hubble tension tend to make the $S_8$ tension worse, mostly due to the reasons described in the previous subsection. In contrast, the baryon clumping generally reduces the value of $S_8$ deduced from CMB, thus also helping with another tension. The primary reason for the lower $S_8$ (and $\Omega_m$) values is that the best fit $\Omega_mh^2$ value in the clumping model is largely the same as in \lcdm, while $h$ is increased. Fig.~\ref{fig:s8} compares the $S_8$-$\Omega_m$ joint posteriors in the Planck+BAO+SN best-fit \lcdm\ model to those in the M1 \lcdmb, together with the DES Year 1 contours.

Based on preliminary results from MHD simulations described in the next section, it is clear that one should not attribute too much weight to observational constraints derived from the three-zone model. However, for general reasons discussed in Sec.~\ref{sec:bao}, it is clear that baryon clumping could at best raise the CMB-based value of the Hubble constant to $H_0 \sim 70$ km/s/Mpc. This may or may not turn out to be enough but, even if the $H_0$ tension was not fully relieved, a detection of clumping would be highly interesting by itself, as it would provide (indirect) evidence of the PMF. If no clumping is detected, it would provide the tightest constraint on the PMF strength.

\section{What to expect from MHD simulations}
\label{sec:mhd}

The three-zone model may capture much of the physics of baryon clumping due to a PMF before recombination. It is not, however, sufficiently accurate for comparison to current and future high precision CMB data as subtle details are neglected or simply assumed. The M1 three-zone model, employed in earlier sections, was originally introduced in~\cite{Jedamzik:2013gua} to derive the preliminary stringent limit on the PMF strength from the (older) CMB data at that time. It assumed a particular baryon PDF and neglected its evolution. Furthermore, at the time of publication of \cite{Jedamzik:2013gua}, not all of the  effects that impact the ionization history in the presence of PMFs were known.  Since then, it became apparent that a complete exhaustive analysis, employing numerical MHD simulations and Monte Carlo methods for radiation transfer is required, which is currently underway~\cite{KJ_TA_2023}.

\begin{figure*}[tbph!]
  \centering
    \includegraphics[width=0.495\textwidth]{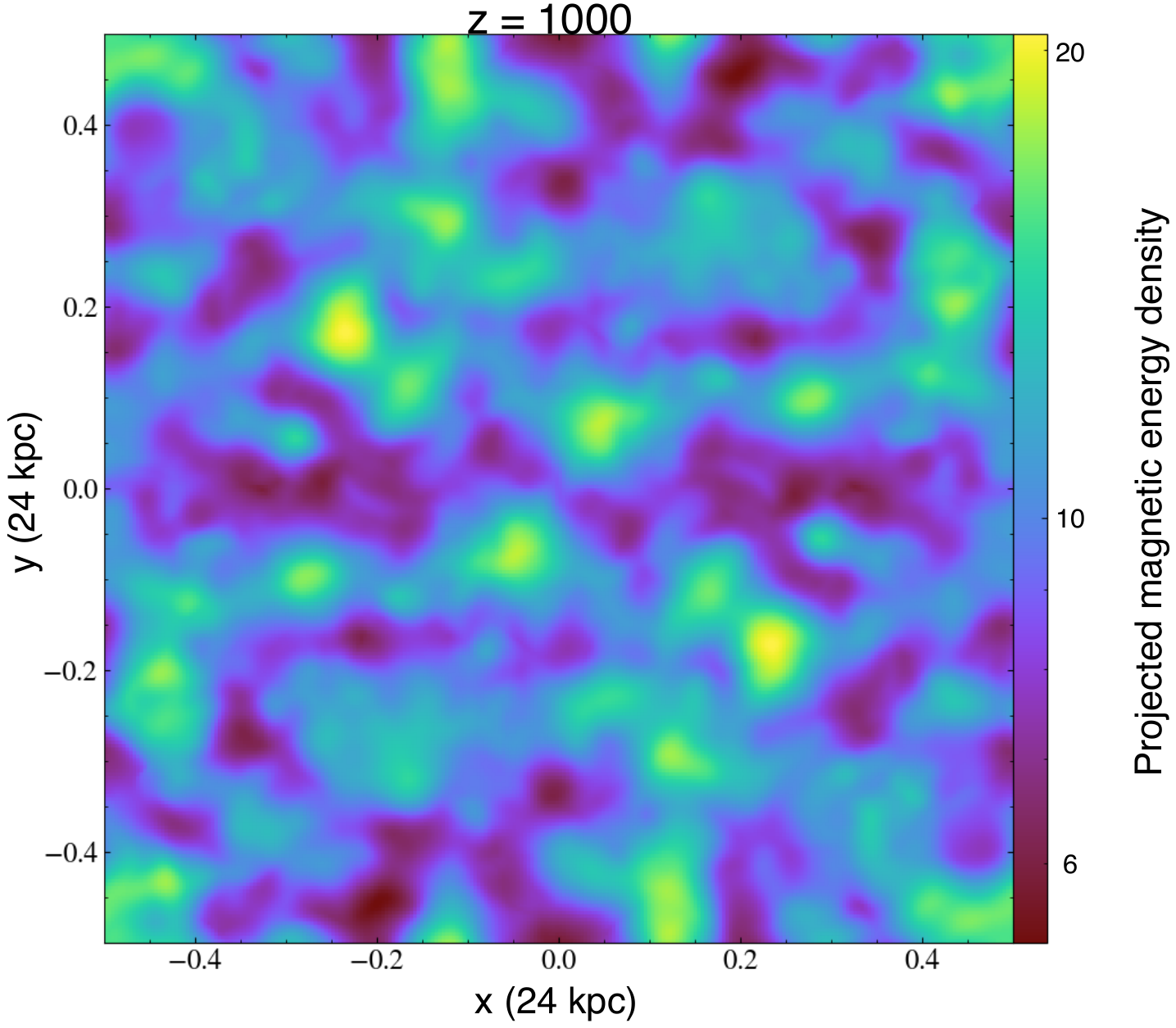}
    \includegraphics[width=0.495\textwidth]{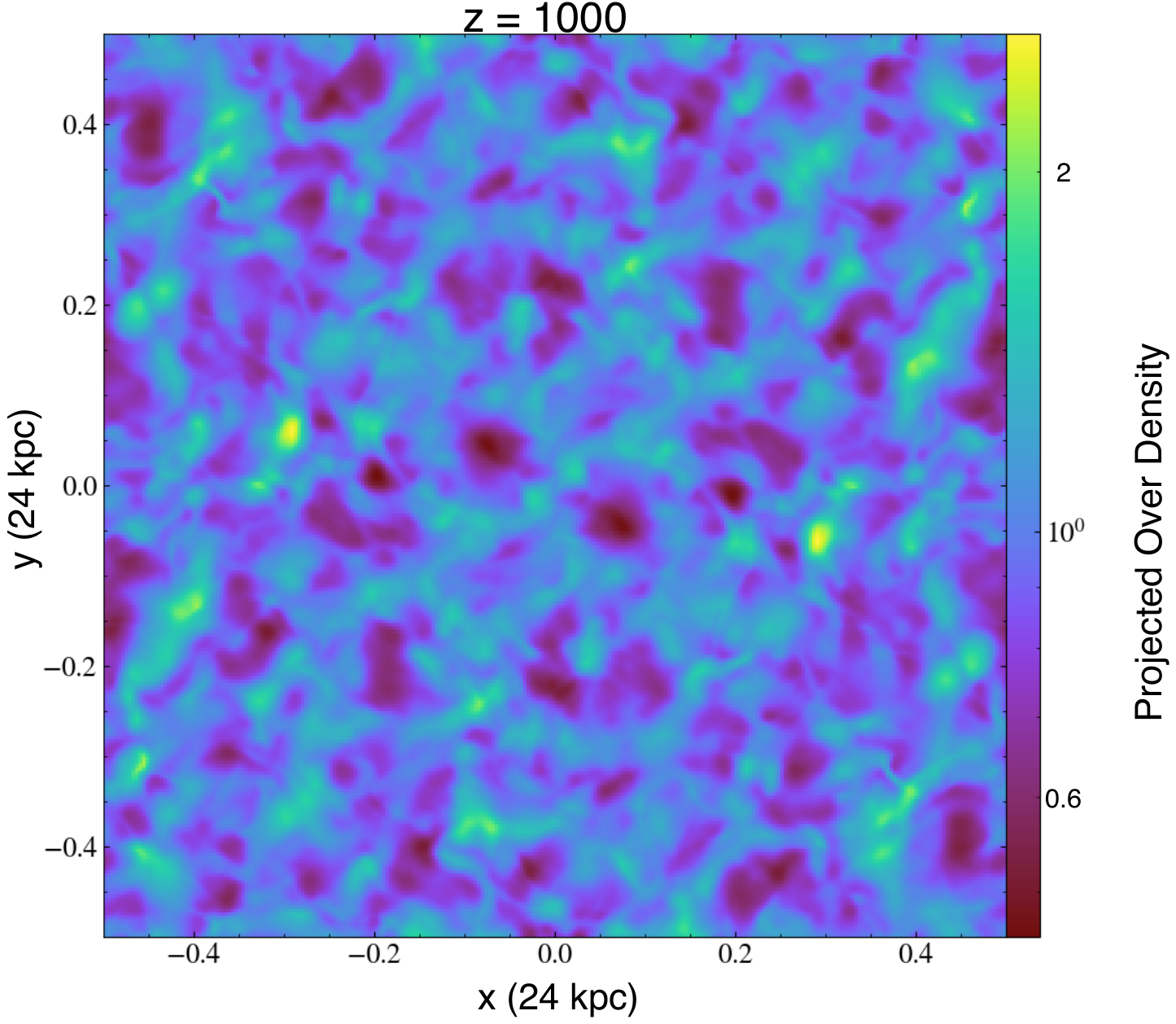}
  \caption{  \label{fig:mhd}
Projected magnetic energy density (left panel) and baryon overdensity (right panel) in a MHD simulation of a stochastic non-helical magnetic field with a Batchelor spectrum at redshift $z = 1000$. The initial conditions, set at $z = 4500$, were uniform baryon density, vanishing peculiar motions and a magnetic field
with a root-mean-square amplitude of $0.53$ nG, or $c_{A,{\rm rms}} = 12 c_s$.
}
\end{figure*}

Fig.~\ref{fig:mhd} shows results of a recent numerical MHD simulation~\cite{KJ_TA_2023}. The left panel shows the projected magnetic energy density and the right panel shows the projected baryon overdensity $\rho/\langle\rho\rangle$ from a $256^3$ simulation of a $(24 {\rm kpc})^3$ volume permeated by a stochastic non-helical magnetic field at redshift $z = 1000$. The initial conditions were a homogeneous baryon density, vanishing peculiar velocities, and a field of root-mean-square amplitude $B_{\rm rms} = 0.53$ nG (corresponding to $c_{A,{\rm rms}} = 12 c_s$) at $z = 4500$. One can see that substantial baryon density fluctuations $\delta\rho /\rho \sim 1$ on $\sim$ kpc scales are generated by redshift $z = 1000$. Most of the volume is occupied by underdense regions, whereas substantial overdensities exist in small
pockets. A small volume fraction of $\sim 10^{-3}$ has overdensities above $10$, that are not visible in the figure as it shows the projected density. In comparison, the magnetic field energy density is much more diffuse. The maximum clumping factor $b\approx 1.5$ is attained around $z \approx 1000$. The corresponding baryon PDF is very skew-positive, {\it i.e.} has a substantial positive third moment, and is significantly different than the PDF assumed in M1. At redshift $z = 1000$, the magnetic field has dissipated to $B_{\rm rms} = 0.2$ nG, whereas the ``final'' field is $B_{\rm rms} = 0.044$ nG at $z \approx 10$.

\begin{figure}[tbph!]
\includegraphics[angle=0,width=0.95\textwidth]{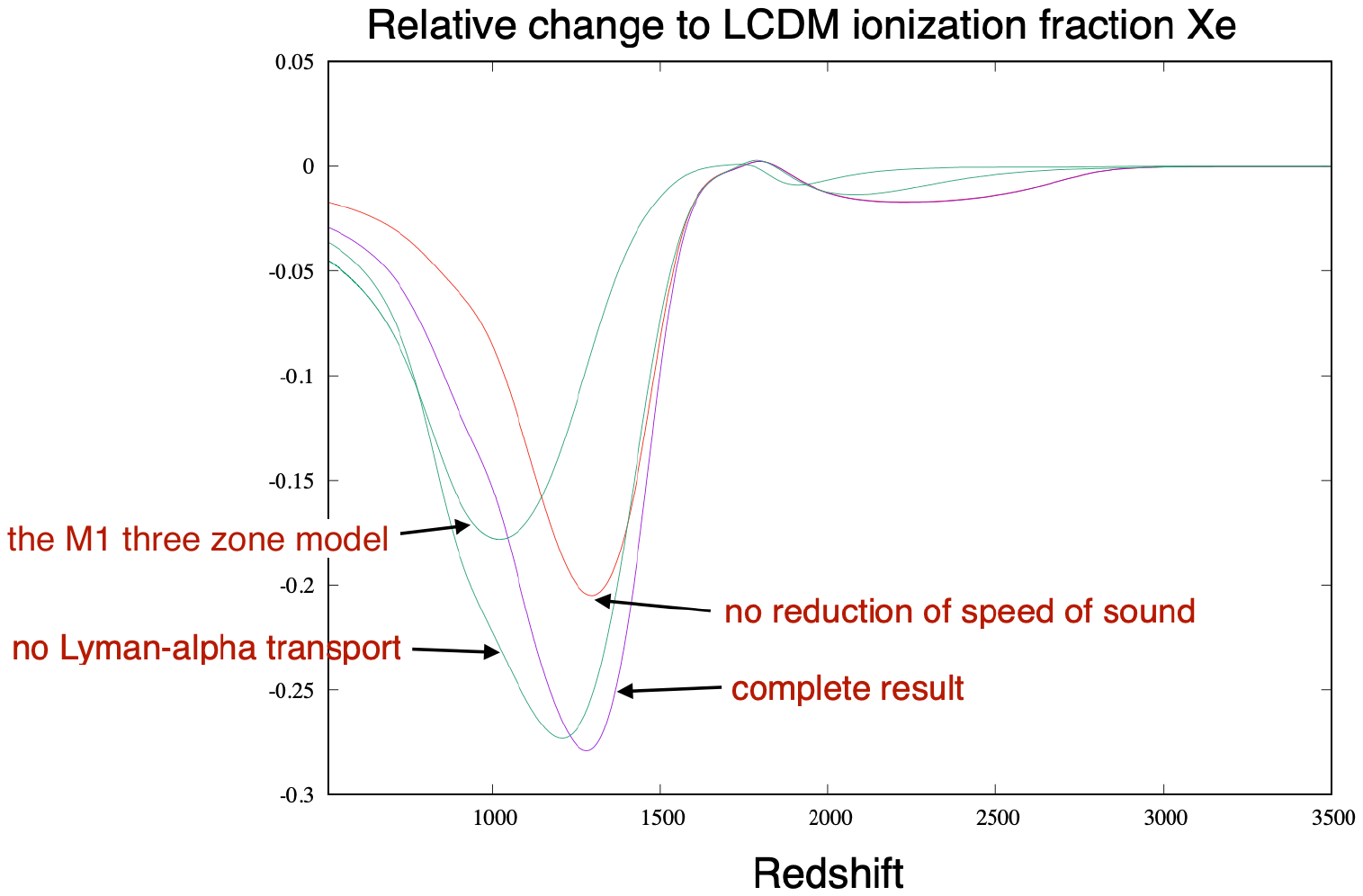}
\caption{Relative change in the average ionization fraction $\Delta X_e$ compared to \lcdm\  
for numerical MHD simulations with (a) all known effects taken into account, (b) neglecting the Lyman-$\alpha$ transport and, (c) neglecting the reduction of the speed of sound during recombination, as labeled. A relatively strong magnetic field with scale-invariant spectrum was assumed. The prediction of the M1 model with $b = 0.5$ is shown for comparison.}
\label{fig:relative}
\end{figure}

The study in \cite{KJ_TA_2023} established several new insights:
\begin{itemize}
\item Even knowing the full evolution of the baryon PDF is not sufficient for an accurate computation of the ionization history. Rather, the density evolution of each fluid element needs to be known. 
\item The clumping factor receives large contributions from rare very high density regions. These, however, do not contribute much to the bulk of the recombination. Therefore, the clumping factor $b$ is not a useful parameter for gauging the effect on the ionization history,
\item A fraction of Lyman-$\alpha$ photons emitted during the recombination process locally may actually travel to other regions with different local conditions. The resultant cosmic average ionization is significantly changed due to this Lyman-$\alpha$ transport.
\item The drop in the speed of sound during recombination by a factor of $1/\sqrt{2}$, due the diminishing electron pressure, leads to further baryon compression by the PMF during recombination.  
\end{itemize}
The last two points are illustrated in Fig.~\ref{fig:relative}, showing that all know effects have to be taken into account for a precision computation of the ionization history.

Ultimately, results of such numerical simulations should be used for precision tests against current and future CMB data. A meaningful comparison faces difficulties due to the required immense CPU time. In theory, given the PMF spectrum and helicity, for each magnetic field strength and cosmological parameters such as baryon density, dark matter density, etc., an independent large simulation should be performed. For the large number of cosmological parameters sampled in such tests, it would be impossible. Fortunately, the effects on $X_e$ from changing the cosmological parameters are much smaller than that of the PMF parameters of appreciable amplitude, such that they may be added linearly to a good approximation. Given a particular set of cosmological parameters, numerical simulations still have to be performed for different initial PMF strengths. This can be done for a discrete set of strengths with subsequent interpolation. The numerical simulation have to be of a significant box size to cover a sufficient dynamical range to simulate all the scales which produce density fluctuations and dissipate afterwards between helium and hydrogen recombination, and for the particular realization of the stochastic field ({\it i.e.} the choice of the random number) to result in a small variance in the results. This points to fairly large simulations and requires extensive amounts of CPU time.

\section{Summary}
\label{sec: summary}

PMFs are often categorized as ``exotic physics'' because, on one hand, there is no firm theoretical prediction of their existence within standard models of particle physics and cosmology, and also because the observational evidence supporting the PMF hypothesis, while increasingly compelling, is still indirect and can, at least in principle, be explained via less exotic astrophysical processes. However, PMFs are a rather mature field of research sustained for over half-a-century by a small but capable community of researchers. The reason for this longevity is not (only) due to PMFs often playing the role of a ``ghost fairy'' (using the term much liked by a certain famous cosmologist) coming to the rescue whenever there is a new unexplained phenomenon. Rather, most of the researchers working on the PMFs are inspired by the fact that their existence in the early universe is inevitable. It is not a question of whether they were generated at some level, {\it e.g.} during the known phase transitions, but whether they would be of sufficient strength to make a detectable impact. If there is even a remote chance of detecting them, it is worth trying, as they would provide an invaluable insight and a new window into the physics of the early universe.

It is the above context that, in our view, distinguishes the PMF proposal for relieving the Hubble tension from others. It is not a "fairy" explanation invented to solve the puzzle of the day, but rather an observation that if the primordial universe was magnetized at a level sufficient to be relevant to the magnetic fields we see in galaxies and clusters, it could also help with alleviating the $H_0$ problem. This is a fully falsifiable proposal -- there is no new physics to invent when it comes to working out the details of recombination in the presence of a PMF. The required MHD simulations are challenging but doable, and there will be a firm prediction in a reasonable time that can then be tested against the data.

 Like other models that aim to relieve the Hubble tension by lowering the sound horizon at recombination, baryon clumping can not raise the value of $H_0$ beyond $\sim$70 km/s/Mpc, which is still over $2\sigma$ lower than the SH0ES measurement. However, even if PMFs do not fully resolve the Hubble tension, finding evidence for them in the CMB would be a major discovery in its own right. Alternatively, if not detected, accounting for their impact on recombination when analyzing data from future CMB experiments, such as Simons Observatory and CMB-S4, would lead to the tightest constraints on a PMF.

\acknowledgments

We thank Tom Abel, Lennart Balkenhol, Silvia Galli, Tanmay Vachaspati and Gong-Bo Zhao for collaborations and discussions. This research was enabled in part by support provided by the BC DRI Group and the Digital Research Alliance of Canada (\url{alliancecan.ca}), and the National Sciences and Engineering Research Council of Canada.


\bibliographystyle{JHEP}
\bibliography{pmf}

\providecommand{\href}[2]{#2}\begingroup\raggedright\begin{thebibliography}{100}

\bibitem{Riess:2021jrx}
A.G.~Riess et~al., \emph{{A Comprehensive Measurement of the Local Value of the
  Hubble Constant with 1 km s$^{?1}$ Mpc$^{?1}$ Uncertainty from the Hubble
  Space Telescope and the SH0ES Team}},
  \href{https://doi.org/10.3847/2041-8213/ac5c5b}{\emph{Astrophys. J. Lett.}
  {\bfseries 934} (2022) L7}
  [\href{https://arxiv.org/abs/2112.04510}{{\ttfamily 2112.04510}}].

\bibitem{Planck:2018vyg}
{\scshape Planck} collaboration, \emph{{Planck 2018 results. VI. Cosmological
  parameters}},
  \href{https://doi.org/10.1051/0004-6361/201833910}{\emph{Astron. Astrophys.}
  {\bfseries 641} (2020) A6}
  [\href{https://arxiv.org/abs/1807.06209}{{\ttfamily 1807.06209}}].

\bibitem{Abdalla:2022yfr}
E.~Abdalla et~al., \emph{{Cosmology intertwined: A review of the particle
  physics, astrophysics, and cosmology associated with the cosmological
  tensions and anomalies}},
  \href{https://doi.org/10.1016/j.jheap.2022.04.002}{\emph{JHEAp} {\bfseries
  34} (2022) 49} [\href{https://arxiv.org/abs/2203.06142}{{\ttfamily
  2203.06142}}].

\bibitem{Pesce:2020xfe}
D.~Pesce et~al., \emph{{The Megamaser Cosmology Project. XIII. Combined Hubble
  constant constraints}},
  \href{https://doi.org/10.3847/2041-8213/ab75f0}{\emph{Astrophys. J.}
  {\bfseries 891} (2020) L1}
  [\href{https://arxiv.org/abs/2001.09213}{{\ttfamily 2001.09213}}].

\bibitem{Wong:2019kwg}
K.C.~Wong et~al., \emph{{H0LiCOW XIII. A 2.4\% measurement of $H_{0}$ from
  lensed quasars: $5.3\sigma$ tension between early and late-Universe probes}},
   \href{https://arxiv.org/abs/1907.04869}{{\ttfamily 1907.04869}}.

\bibitem{Shajib:2019toy}
{\scshape DES} collaboration, \emph{{STRIDES: a 3.9 per cent measurement of the
  Hubble constant from the strong lens system DES J0408\ensuremath{-}5354}},
  \href{https://doi.org/10.1093/mnras/staa828}{\emph{Mon. Not. Roy. Astron.
  Soc.} {\bfseries 494} (2020) 6072}
  [\href{https://arxiv.org/abs/1910.06306}{{\ttfamily 1910.06306}}].

\bibitem{Harvey:2020lwf}
D.~Harvey, \emph{{A 4\% measurement of $H_0$ using the cumulative distribution
  of strong-lensing time delays in doubly-imaged quasars}},
  \href{https://arxiv.org/abs/2011.09488}{{\ttfamily 2011.09488}}.

\bibitem{Millon:2019slk}
M.~Millon et~al., \emph{{TDCOSMO. I. An exploration of systematic uncertainties
  in the inference of $H_0$ from time-delay cosmography}},
  \href{https://doi.org/10.1051/0004-6361/201937351}{\emph{Astron. Astrophys.}
  {\bfseries 639} (2020) A101}
  [\href{https://arxiv.org/abs/1912.08027}{{\ttfamily 1912.08027}}].

\bibitem{Freedman:2020dne}
W.L.~Freedman, B.F.~Madore, T.~Hoyt, I.S.~Jang, R.~Beaton, M.G.~Lee et~al.,
  \emph{{Calibration of the Tip of the Red Giant Branch (TRGB)}},
  \href{https://arxiv.org/abs/2002.01550}{{\ttfamily 2002.01550}}.

\bibitem{Ivanov:2019pdj}
M.M.~Ivanov, M.~Simonovi\'c and M.~Zaldarriaga, \emph{{Cosmological Parameters
  from the BOSS Galaxy Power Spectrum}},
  \href{https://doi.org/10.1088/1475-7516/2020/05/042}{\emph{JCAP} {\bfseries
  05} (2020) 042} [\href{https://arxiv.org/abs/1909.05277}{{\ttfamily
  1909.05277}}].

\bibitem{Aiola:2020azj}
{\scshape ACT} collaboration, \emph{{The Atacama Cosmology Telescope: DR4 Maps
  and Cosmological Parameters}},
  \href{https://doi.org/10.1088/1475-7516/2020/12/047}{\emph{JCAP} {\bfseries
  12} (2020) 047} [\href{https://arxiv.org/abs/2007.07288}{{\ttfamily
  2007.07288}}].

\bibitem{Alam:2020sor}
{\scshape eBOSS} collaboration, \emph{{The Completed SDSS-IV extended Baryon
  Oscillation Spectroscopic Survey: Cosmological Implications from two Decades
  of Spectroscopic Surveys at the Apache Point observatory}},
  \href{https://arxiv.org/abs/2007.08991}{{\ttfamily 2007.08991}}.

\bibitem{Jedamzik:2013gua}
K.~Jedamzik and T.~Abel, \emph{{Small-scale primordial magnetic fields and
  anisotropies in the cosmic microwave background radiation}},
  \href{https://doi.org/10.1088/1475-7516/2013/10/050}{\emph{JCAP} {\bfseries
  1310} (2013) 050} [\href{https://arxiv.org/abs/1108.2517}{{\ttfamily
  1108.2517}}].

\bibitem{Jedamzik:2018itu}
K.~Jedamzik and A.~Saveliev, \emph{{Stringent Limit on Primordial Magnetic
  Fields from the Cosmic Microwave Background Radiation}},
  \href{https://doi.org/10.1103/PhysRevLett.123.021301}{\emph{Phys. Rev. Lett.}
  {\bfseries 123} (2019) 021301}
  [\href{https://arxiv.org/abs/1804.06115}{{\ttfamily 1804.06115}}].

\bibitem{Peebles:1994xt}
P.J.E.~Peebles, \emph{{Principles of physical cosmology}} (1994).

\bibitem{Jedamzik:2020krr}
K.~Jedamzik and L.~Pogosian, \emph{{Relieving the Hubble tension with
  primordial magnetic fields}},
  \href{https://doi.org/10.1103/PhysRevLett.125.181302}{\emph{Phys. Rev. Lett.}
  {\bfseries 125} (2020) 181302}
  [\href{https://arxiv.org/abs/2004.09487}{{\ttfamily 2004.09487}}].

\bibitem{Thiele:2021okz}
L.~Thiele, Y.~Guan, J.C.~Hill, A.~Kosowsky and D.N.~Spergel, \emph{{Can
  small-scale baryon inhomogeneities resolve the Hubble tension? An
  investigation with ACT DR4}},
  \href{https://arxiv.org/abs/2105.03003}{{\ttfamily 2105.03003}}.

\bibitem{Rashkovetskyi:2021}
M.~{Rashkovetskyi}, J.B.~{Mu{\~n}oz}, D.J.~{Eisenstein} and C.~{Dvorkin},
  \emph{{Small-scale Clumping at Recombination and the Hubble Tension}},
  {\emph{arXiv e-prints} (2021) arXiv:2108.02747}
  [\href{https://arxiv.org/abs/2108.02747}{{\ttfamily 2108.02747}}].

\bibitem{Galli:2021mxk}
S.~Galli, L.~Pogosian, K.~Jedamzik and L.~Balkenhol, \emph{{Consistency of
  Planck, ACT, and SPT constraints on magnetically assisted recombination and
  forecasts for future experiments}},
  \href{https://doi.org/10.1103/PhysRevD.105.023513}{\emph{Phys. Rev. D}
  {\bfseries 105} (2022) 023513}
  [\href{https://arxiv.org/abs/2109.03816}{{\ttfamily 2109.03816}}].

\bibitem{SPT-3G:2022hvq}
{\scshape SPT-3G} collaboration, \emph{{A Measurement of the CMB Temperature
  Power Spectrum and Constraints on Cosmology from the SPT-3G 2018 TT/TE/EE
  Data Set}},  \href{https://arxiv.org/abs/2212.05642}{{\ttfamily 2212.05642}}.

\bibitem{KJ_TA_2023}
K.~Jedamzik and T.~Abel, ``{Cosmological recombination in the presence of
  primordial magnetic fields (in preparation), 2023}.''.

\bibitem{Hoyle:1958}
F.~Hoyle{\emph{XI Solvay Congress, Brussells} (1958) }.

\bibitem{Zeldovich:1965}
Y.~Zeldovich, \emph{{Magnetic model of the universe}}, {\emph{J. Exptl.
  Theoret. Phys. (U.S.S.R.)} {\bfseries 48} (1965) 986}.

\bibitem{Widrow:2002ud}
L.M.~Widrow, \emph{{Origin of galactic and extragalactic magnetic fields}},
  \href{https://doi.org/10.1103/RevModPhys.74.775}{\emph{Rev. Mod. Phys.}
  {\bfseries 74} (2002) 775}
  [\href{https://arxiv.org/abs/astro-ph/0207240}{{\ttfamily
  astro-ph/0207240}}].

\bibitem{Widrow:2011hs}
L.M.~Widrow, D.~Ryu, D.R.G.~Schleicher, K.~Subramanian, C.G.~Tsagas and
  R.A.~Treumann, \emph{{The First Magnetic Fields}},
  \href{https://doi.org/10.1007/s11214-011-9833-5}{\emph{Space Sci. Rev.}
  {\bfseries 166} (2012) 37} [\href{https://arxiv.org/abs/1109.4052}{{\ttfamily
  1109.4052}}].

\bibitem{Vachaspati:2020blt}
T.~Vachaspati, \emph{{Progress on Cosmological Magnetic Fields}},
  \href{https://arxiv.org/abs/2010.10525}{{\ttfamily 2010.10525}}.

\bibitem{Brandenburg:2004jv}
A.~Brandenburg and K.~Subramanian, \emph{{Astrophysical magnetic fields and
  nonlinear dynamo theory}},
  \href{https://doi.org/10.1016/j.physrep.2005.06.005}{\emph{Phys. Rept.}
  {\bfseries 417} (2005) 1}
  [\href{https://arxiv.org/abs/astro-ph/0405052}{{\ttfamily
  astro-ph/0405052}}].

\bibitem{Athreya:1998}
R.M.~{Athreya}, V.K.~{Kapahi}, P.J.~{McCarthy} and W.~{van Breugel},
  \emph{{Large rotation measures in radio galaxies at Z > 2}}, {\emph{Astron.
  Astrophys.} {\bfseries 329} (1998) 809}.

\bibitem{Beck:2013gca}
A.M.~Beck, K.~Dolag, H.~Lesch and P.P.~Kronberg, \emph{{Strong magnetic fields
  and large rotation measures in protogalaxies by supernova seeding}},
  \href{https://doi.org/10.1093/mnras/stt1549}{\emph{Mon. Not. Roy. Astron.
  Soc.} {\bfseries 435} (2013) 3575}
  [\href{https://arxiv.org/abs/1308.3440}{{\ttfamily 1308.3440}}].

\bibitem{DSeifried:2013zeu}
D.~Seifried, R.~Banerjee and D.~Schleicher, \emph{{Supernova explosions in
  magnetized, primordial dark matter haloes}},
  \href{https://doi.org/10.1093/mnras/stu294}{\emph{Mon. Not. Roy. Astron.
  Soc.} {\bfseries 440} (2014) 24}
  [\href{https://arxiv.org/abs/1311.4991}{{\ttfamily 1311.4991}}].

\bibitem{Vachaspati:1991nm}
T.~Vachaspati, \emph{{Magnetic fields from cosmological phase transitions}},
  \href{https://doi.org/10.1016/0370-2693(91)90051-Q}{\emph{Phys. Lett.}
  {\bfseries B265} (1991) 258}.

\bibitem{Turner:1987bw}
M.S.~Turner and L.M.~Widrow, \emph{{Inflation Produced, Large Scale Magnetic
  Fields}}, \href{https://doi.org/10.1103/PhysRevD.37.2743}{\emph{Phys. Rev.}
  {\bfseries D37} (1988) 2743}.

\bibitem{Ratra:1991bn}
B.~Ratra, \emph{{Cosmological 'seed' magnetic field from inflation}},
  {\emph{Astrophys. J.} {\bfseries 391} (1992) L1}.

\bibitem{Durrer:2013pga}
R.~Durrer and A.~Neronov, \emph{{Cosmological Magnetic Fields: Their
  Generation, Evolution and Observation}},
  \href{https://doi.org/10.1007/s00159-013-0062-7}{\emph{Astron.Astrophys.Rev.}
  {\bfseries 21} (2013) 62} [\href{https://arxiv.org/abs/1303.7121}{{\ttfamily
  1303.7121}}].

\bibitem{Subramanian:2015lua}
K.~Subramanian, \emph{{The origin, evolution and signatures of primordial
  magnetic fields}},
  \href{https://doi.org/10.1088/0034-4885/79/7/076901}{\emph{Rept. Prog. Phys.}
  {\bfseries 79} (2016) 076901}
  [\href{https://arxiv.org/abs/1504.02311}{{\ttfamily 1504.02311}}].

\bibitem{Jedamzik:1996wp}
K.~Jedamzik, V.~Katalinic and A.V.~Olinto, \emph{{Damping of cosmic magnetic
  fields}}, \href{https://doi.org/10.1103/PhysRevD.57.3264}{\emph{Phys. Rev. D}
  {\bfseries 57} (1998) 3264}
  [\href{https://arxiv.org/abs/astro-ph/9606080}{{\ttfamily
  astro-ph/9606080}}].

\bibitem{Subramanian:1997gi}
K.~Subramanian and J.D.~Barrow, \emph{{Magnetohydrodynamics in the early
  universe and the damping of noninear Alfven waves}},
  \href{https://doi.org/10.1103/PhysRevD.58.083502}{\emph{Phys. Rev.}
  {\bfseries D58} (1998) 083502}
  [\href{https://arxiv.org/abs/astro-ph/9712083}{{\ttfamily
  astro-ph/9712083}}].

\bibitem{Neronov:1900zz}
A.~Neronov and I.~Vovk, \emph{{Evidence for strong extragalactic magnetic
  fields from Fermi observations of TeV blazars}},
  \href{https://doi.org/10.1126/science.1184192}{\emph{Science} {\bfseries 328}
  (2010) 73} [\href{https://arxiv.org/abs/1006.3504}{{\ttfamily 1006.3504}}].

\bibitem{Tavecchio:2010mk}
F.~Tavecchio, G.~Ghisellini, L.~Foschini, G.~Bonnoli, G.~Ghirlanda and
  P.~Coppi, \emph{{The intergalactic magnetic field constrained by Fermi/LAT
  observations of the TeV blazar 1ES 0229+200}},
  \href{https://doi.org/10.1111/j.1745-3933.2010.00884.x}{\emph{Mon. Not. Roy.
  Astron. Soc.} {\bfseries 406} (2010) L70}
  [\href{https://arxiv.org/abs/1004.1329}{{\ttfamily 1004.1329}}].

\bibitem{Taylor:2011bn}
A.~Taylor, I.~Vovk and A.~Neronov, \emph{{Extragalactic magnetic fields
  constraints from simultaneous GeV-TeV observations of blazars}},
  {\emph{Astron.Astrophys.} {\bfseries 529} (2011) A144}
  [\href{https://arxiv.org/abs/1101.0932}{{\ttfamily 1101.0932}}].

\bibitem{Dolag:2010ni}
K.~Dolag, M.~Kachelriess, S.~Ostapchenko and R.~Tomas, \emph{{Lower limit on
  the strength and filling factor of extragalactic magnetic fields}},
  \href{https://doi.org/10.1088/2041-8205/727/1/L4}{\emph{Astrophys. J. Lett.}
  {\bfseries 727} (2011) L4} [\href{https://arxiv.org/abs/1009.1782}{{\ttfamily
  1009.1782}}].

\bibitem{Broderick:2011av}
A.E.~Broderick, P.~Chang and C.~Pfrommer, \emph{{The Cosmological Impact of
  Luminous TeV Blazars I: Implications of Plasma Instabilities for the
  Intergalactic Magnetic Field and Extragalactic Gamma-Ray Background}},
  \href{https://doi.org/10.1088/0004-637X/752/1/22}{\emph{Astrophys. J.}
  {\bfseries 752} (2012) 22} [\href{https://arxiv.org/abs/1106.5494}{{\ttfamily
  1106.5494}}].

\bibitem{Govoni_2019}
F.~Govoni, E.~Orr{\`{u} }, A.~Bonafede, M.~Iacobelli, R.~Paladino, F.~Vazza
  et~al., \emph{A radio ridge connecting two galaxy clusters in a filament of
  the cosmic web},
  \href{https://doi.org/10.1126/science.aat7500}{\emph{Science} {\bfseries 364}
  (2019) 981}.

\bibitem{CTAConsortium:2018tzg}
{\scshape CTA Consortium} collaboration, B.S.~Acharya et~al., \emph{{Science
  with the Cherenkov Telescope Array}}, WSP (11, 2018),
  \href{https://doi.org/10.1142/10986}{10.1142/10986},
  [\href{https://arxiv.org/abs/1709.07997}{{\ttfamily 1709.07997}}].

\bibitem{Korochkin:2020pvg}
A.~Korochkin, O.~Kalashev, A.~Neronov and D.~Semikoz, \emph{{Sensitivity reach
  of gamma-ray measurements for strong cosmological magnetic fields}},
  \href{https://doi.org/10.3847/1538-4357/abc697}{\emph{Astrophys. J.}
  {\bfseries 906} (2021) 116}
  [\href{https://arxiv.org/abs/2007.14331}{{\ttfamily 2007.14331}}].

\bibitem{Subramanian:1998fn}
K.~Subramanian and J.D.~Barrow, \emph{{Microwave background signals from
  tangled magnetic fields}},
  \href{https://doi.org/10.1103/PhysRevLett.81.3575}{\emph{Phys. Rev. Lett.}
  {\bfseries 81} (1998) 3575}
  [\href{https://arxiv.org/abs/astro-ph/9803261}{{\ttfamily
  astro-ph/9803261}}].

\bibitem{Subramanian:2002nh}
K.~Subramanian and J.D.~Barrow, \emph{{Small-scale microwave background
  anisotropies due to tangled primordial magnetic fields}},
  \href{https://doi.org/10.1046/j.1365-8711.2002.05854.x}{\emph{Mon. Not. Roy.
  Astron. Soc.} {\bfseries 335} (2002) L57}
  [\href{https://arxiv.org/abs/astro-ph/0205312}{{\ttfamily
  astro-ph/0205312}}].

\bibitem{Mack:2001gc}
A.~Mack, T.~Kahniashvili and A.~Kosowsky, \emph{{Microwave background
  signatures of a primordial stochastic magnetic field}},
  \href{https://doi.org/10.1103/PhysRevD.65.123004}{\emph{Phys. Rev.}
  {\bfseries D65} (2002) 123004}
  [\href{https://arxiv.org/abs/astro-ph/0105504}{{\ttfamily
  astro-ph/0105504}}].

\bibitem{Lewis:2004kg}
A.~Lewis, \emph{{Observable primordial vector modes}},
  \href{https://doi.org/10.1103/PhysRevD.70.043518}{\emph{Phys. Rev.}
  {\bfseries D70} (2004) 043518}
  [\href{https://arxiv.org/abs/astro-ph/0403583}{{\ttfamily
  astro-ph/0403583}}].

\bibitem{Kahniashvili:2005xe}
T.~Kahniashvili and B.~Ratra, \emph{{Effects of cosmological magnetic helicity
  on the cosmic microwave background}},
  \href{https://doi.org/10.1103/PhysRevD.71.103006}{\emph{Phys.Rev.} {\bfseries
  D71} (2005) 103006} [\href{https://arxiv.org/abs/astro-ph/0503709}{{\ttfamily
  astro-ph/0503709}}].

\bibitem{Chen:2004nf}
G.~Chen, P.~Mukherjee, T.~Kahniashvili, B.~Ratra and Y.~Wang, \emph{{Looking
  for cosmological Alfven waves in WMAP data}},
  \href{https://doi.org/10.1086/422213}{\emph{Astrophys. J.} {\bfseries 611}
  (2004) 655} [\href{https://arxiv.org/abs/astro-ph/0403695}{{\ttfamily
  astro-ph/0403695}}].

\bibitem{Lewis:2004ef}
A.~Lewis, \emph{{CMB anisotropies from primordial inhomogeneous magnetic
  fields}}, \href{https://doi.org/10.1103/PhysRevD.70.043011}{\emph{Phys.Rev.}
  {\bfseries D70} (2004) 043011}
  [\href{https://arxiv.org/abs/astro-ph/0406096}{{\ttfamily
  astro-ph/0406096}}].

\bibitem{Tashiro:2005hc}
H.~Tashiro, N.~Sugiyama and R.~Banerjee, \emph{{Nonlinear evolution of cosmic
  magnetic fields and cosmic microwave background anisotropies}},
  \href{https://doi.org/10.1103/PhysRevD.73.023002}{\emph{Phys. Rev.}
  {\bfseries D73} (2006) 023002}
  [\href{https://arxiv.org/abs/astro-ph/0509220}{{\ttfamily
  astro-ph/0509220}}].

\bibitem{Yamazaki:2006bq}
D.~Yamazaki, K.~Ichiki, T.~Kajino and G.J.~Mathews, \emph{{Constraints on the
  evolution of the primordial magnetic field from the small scale cmb angular
  anisotropy}}, \href{https://doi.org/10.1086/505135}{\emph{Astrophys. J.}
  {\bfseries 646} (2006) 719}
  [\href{https://arxiv.org/abs/astro-ph/0602224}{{\ttfamily
  astro-ph/0602224}}].

\bibitem{Giovannini:2006gz}
M.~Giovannini, \emph{{Entropy perturbations and large-scale magnetic fields}},
  \href{https://doi.org/10.1088/0264-9381/23/15/017}{\emph{Class. Quant. Grav.}
  {\bfseries 23} (2006) 4991}
  [\href{https://arxiv.org/abs/astro-ph/0604134}{{\ttfamily
  astro-ph/0604134}}].

\bibitem{Kahniashvili:2006hy}
T.~Kahniashvili and B.~Ratra, \emph{{CMB anisotropies due to cosmological
  magnetosonic waves}},
  \href{https://doi.org/10.1103/PhysRevD.75.023002}{\emph{Phys. Rev.}
  {\bfseries D75} (2007) 023002}
  [\href{https://arxiv.org/abs/astro-ph/0611247}{{\ttfamily
  astro-ph/0611247}}].

\bibitem{Giovannini:2007qn}
M.~Giovannini and K.E.~Kunze, \emph{{Magnetized CMB observables: A Dedicated
  numerical approach}},
  \href{https://doi.org/10.1103/PhysRevD.77.063003}{\emph{Phys. Rev.}
  {\bfseries D77} (2008) 063003}
  [\href{https://arxiv.org/abs/0712.3483}{{\ttfamily 0712.3483}}].

\bibitem{Yamazaki:2010nf}
D.G.~Yamazaki, K.~Ichiki, T.~Kajino and G.J.~Mathews, \emph{{New Constraints on
  the Primordial Magnetic Field}},
  \href{https://doi.org/10.1103/PhysRevD.81.023008}{\emph{Phys. Rev.}
  {\bfseries D81} (2010) 023008}
  [\href{https://arxiv.org/abs/1001.2012}{{\ttfamily 1001.2012}}].

\bibitem{Paoletti:2010rx}
D.~Paoletti and F.~Finelli, \emph{{CMB Constraints on a Stochastic Background
  of Primordial Magnetic Fields}},
  \href{https://doi.org/10.1103/PhysRevD.83.123533}{\emph{Phys.Rev.} {\bfseries
  D83} (2011) 123533} [\href{https://arxiv.org/abs/1005.0148}{{\ttfamily
  1005.0148}}].

\bibitem{Shaw:2010ea}
J.R.~Shaw and A.~Lewis, \emph{{Constraining Primordial Magnetism}},
  \href{https://doi.org/10.1103/PhysRevD.86.043510}{\emph{Phys. Rev.}
  {\bfseries D86} (2012) 043510}
  [\href{https://arxiv.org/abs/1006.4242}{{\ttfamily 1006.4242}}].

\bibitem{Kunze:2010ys}
K.E.~Kunze, \emph{{CMB anisotropies in the presence of a stochastic magnetic
  field}}, \href{https://doi.org/10.1103/PhysRevD.83.023006}{\emph{Phys. Rev.}
  {\bfseries D83} (2011) 023006}
  [\href{https://arxiv.org/abs/1007.3163}{{\ttfamily 1007.3163}}].

\bibitem{Paoletti:2012bb}
D.~Paoletti and F.~Finelli, \emph{{Constraints on a Stochastic Background of
  Primordial Magnetic Fields with WMAP and South Pole Telescope data}},
  \href{https://doi.org/10.1016/j.physletb.2013.08.065}{\emph{Phys. Lett.}
  {\bfseries B726} (2013) 45}
  [\href{https://arxiv.org/abs/1208.2625}{{\ttfamily 1208.2625}}].

\bibitem{Ballardini:2014jta}
M.~Ballardini, F.~Finelli and D.~Paoletti, \emph{{CMB anisotropies generated by
  a stochastic background of primordial magnetic fields with non-zero
  helicity}}, \href{https://doi.org/10.1088/1475-7516/2015/10/031}{\emph{JCAP}
  {\bfseries 1510} (2015) 031}
  [\href{https://arxiv.org/abs/1412.1836}{{\ttfamily 1412.1836}}].

\bibitem{Ade:2015cva}
{\scshape Planck} collaboration, \emph{{Planck 2015 results. XIX. Constraints
  on primordial magnetic fields}},
  \href{https://doi.org/10.1051/0004-6361/201525821}{\emph{Astron. Astrophys.}
  {\bfseries 594} (2016) A19}
  [\href{https://arxiv.org/abs/1502.01594}{{\ttfamily 1502.01594}}].

\bibitem{Sutton:2017jgr}
D.R.~Sutton, C.~Feng and C.L.~Reichardt, \emph{{Current and Future Constraints
  on Primordial Magnetic Fields}},
  \href{https://doi.org/10.3847/1538-4357/aa85e2}{\emph{Astrophys. J.}
  {\bfseries 846} (2017) 164}
  [\href{https://arxiv.org/abs/1702.01871}{{\ttfamily 1702.01871}}].

\bibitem{Zucca:2016iur}
A.~Zucca, Y.~Li and L.~Pogosian, \emph{{Constraints on Primordial Magnetic
  Fields from Planck combined with the South Pole Telescope CMB B-mode
  polarization measurements}},
  \href{https://doi.org/10.1103/PhysRevD.95.063506}{\emph{Phys. Rev.}
  {\bfseries D95} (2017) 063506}
  [\href{https://arxiv.org/abs/1611.00757}{{\ttfamily 1611.00757}}].

\bibitem{Jedamzik:1999bm}
K.~Jedamzik, V.~Katalinic and A.V.~Olinto, \emph{{A Limit on primordial small
  scale magnetic fields from CMB distortions}},
  \href{https://doi.org/10.1103/PhysRevLett.85.700}{\emph{Phys. Rev. Lett.}
  {\bfseries 85} (2000) 700}
  [\href{https://arxiv.org/abs/astro-ph/9911100}{{\ttfamily
  astro-ph/9911100}}].

\bibitem{Zizzo:2005az}
A.~Zizzo and C.~Burigana, \emph{{On the effect of cyclotron emission on the
  spectral distortions of the cosmic microwave background}},
  \href{https://doi.org/10.1016/j.newast.2005.05.003}{\emph{New Astron.}
  {\bfseries 11} (2005) 1}
  [\href{https://arxiv.org/abs/astro-ph/0505259}{{\ttfamily
  astro-ph/0505259}}].

\bibitem{Kunze:2013uja}
K.E.~Kunze and E.~Komatsu, \emph{{Constraining primordial magnetic fields with
  distortions of the black-body spectrum of the cosmic microwave background:
  pre- and post-decoupling contributions}},
  \href{https://doi.org/10.1088/1475-7516/2014/01/009}{\emph{JCAP} {\bfseries
  1401} (2014) 009} [\href{https://arxiv.org/abs/1309.7994}{{\ttfamily
  1309.7994}}].

\bibitem{Ganc:2014wia}
J.~Ganc and M.S.~Sloth, \emph{{Probing correlations of early magnetic fields
  using mu-distortion}},
  \href{https://doi.org/10.1088/1475-7516/2014/08/018}{\emph{JCAP} {\bfseries
  1408} (2014) 018} [\href{https://arxiv.org/abs/1404.5957}{{\ttfamily
  1404.5957}}].

\bibitem{Paoletti:2018uic}
D.~Paoletti, J.~Chluba, F.~Finelli and J.A.~Rubino-Martin, \emph{{Improved CMB
  anisotropy constraints on primordial magnetic fields from the
  post-recombination ionization history}},
  \href{https://doi.org/10.1093/mnras/sty3521}{\emph{Mon. Not. Roy. Astron.
  Soc.} {\bfseries 484} (2019) 185}
  [\href{https://arxiv.org/abs/1806.06830}{{\ttfamily 1806.06830}}].

\bibitem{Sethi:2004pe}
S.K.~Sethi and K.~Subramanian, \emph{{Primordial magnetic fields in the
  post-recombination era and early reionization}},
  \href{https://doi.org/10.1111/j.1365-2966.2004.08520.x}{\emph{Mon. Not. Roy.
  Astron. Soc.} {\bfseries 356} (2005) 778}
  [\href{https://arxiv.org/abs/astro-ph/0405413}{{\ttfamily
  astro-ph/0405413}}].

\bibitem{Kunze:2014eka}
K.E.~Kunze and E.~Komatsu, \emph{{Constraints on primordial magnetic fields
  from the optical depth of the cosmic microwave background}},
  \href{https://doi.org/10.1088/1475-7516/2015/06/027}{\emph{JCAP} {\bfseries
  1506} (2015) 027} [\href{https://arxiv.org/abs/1501.00142}{{\ttfamily
  1501.00142}}].

\bibitem{Chluba:2015lpa}
J.~Chluba, D.~Paoletti, F.~Finelli and J.-A.~Rubi\~no Mart\'\i{}n,
  \emph{{Effect of primordial magnetic fields on the ionization history}},
  \href{https://doi.org/10.1093/mnras/stv1096}{\emph{Mon. Not. Roy. Astron.
  Soc.} {\bfseries 451} (2015) 2244}
  [\href{https://arxiv.org/abs/1503.04827}{{\ttfamily 1503.04827}}].

\bibitem{Durrer:1999bk}
R.~Durrer, P.G.~Ferreira and T.~Kahniashvili, \emph{{Tensor microwave
  anisotropies from a stochastic magnetic field}},
  \href{https://doi.org/10.1103/PhysRevD.61.043001}{\emph{Phys. Rev.}
  {\bfseries D61} (2000) 043001}
  [\href{https://arxiv.org/abs/astro-ph/9911040}{{\ttfamily
  astro-ph/9911040}}].

\bibitem{Seshadri:2000ky}
T.R.~Seshadri and K.~Subramanian, \emph{{CMBR polarization signals from tangled
  magnetic fields}},
  \href{https://doi.org/10.1103/PhysRevLett.87.101301}{\emph{Phys. Rev. Lett.}
  {\bfseries 87} (2001) 101301}
  [\href{https://arxiv.org/abs/astro-ph/0012056}{{\ttfamily
  astro-ph/0012056}}].

\bibitem{Subramanian:2003sh}
K.~Subramanian, T.R.~Seshadri and J.~Barrow, \emph{{Small - scale CMB
  polarization anisotropies due to tangled primordial magnetic fields}},
  \href{https://doi.org/10.1046/j.1365-8711.2003.07003.x}{\emph{Mon. Not. Roy.
  Astron. Soc.} {\bfseries 344} (2003) L31}
  [\href{https://arxiv.org/abs/astro-ph/0303014}{{\ttfamily
  astro-ph/0303014}}].

\bibitem{Mollerach:2003nq}
S.~Mollerach, D.~Harari and S.~Matarrese, \emph{{CMB polarization from
  secondary vector and tensor modes}},
  \href{https://doi.org/10.1103/PhysRevD.69.063002}{\emph{Phys. Rev.}
  {\bfseries D69} (2004) 063002}
  [\href{https://arxiv.org/abs/astro-ph/0310711}{{\ttfamily
  astro-ph/0310711}}].

\bibitem{Scoccola:2004ke}
C.~Scoccola, D.~Harari and S.~Mollerach, \emph{{B polarization of the CMB from
  Faraday rotation}},
  \href{https://doi.org/10.1103/PhysRevD.70.063003}{\emph{Phys. Rev.}
  {\bfseries D70} (2004) 063003}
  [\href{https://arxiv.org/abs/astro-ph/0405396}{{\ttfamily
  astro-ph/0405396}}].

\bibitem{Kosowsky:2004zh}
A.~Kosowsky, T.~Kahniashvili, G.~Lavrelashvili and B.~Ratra, \emph{{Faraday
  rotation of the Cosmic Microwave Background polarization by a stochastic
  magnetic field}},
  \href{https://doi.org/10.1103/PhysRevD.71.043006}{\emph{Phys. Rev.}
  {\bfseries D71} (2005) 043006}
  [\href{https://arxiv.org/abs/astro-ph/0409767}{{\ttfamily
  astro-ph/0409767}}].

\bibitem{Pogosian:2012jd}
L.~Pogosian, T.~Vachaspati and A.~Yadav, \emph{{Primordial Magnetism in CMB
  B-modes}}, \href{https://doi.org/10.1139/cjp-2012-0401}{\emph{Can. J. Phys.}
  {\bfseries 91} (2013) 451} [\href{https://arxiv.org/abs/1210.0308}{{\ttfamily
  1210.0308}}].

\bibitem{Kahniashvili:2014dfa}
T.~Kahniashvili, Y.~Maravin, G.~Lavrelashvili and A.~Kosowsky,
  \emph{{Primordial Magnetic Helicity Constraints from WMAP Nine-Year Data}},
  \href{https://doi.org/10.1103/PhysRevD.90.083004}{\emph{Phys. Rev.}
  {\bfseries D90} (2014) 083004}
  [\href{https://arxiv.org/abs/1408.0351}{{\ttfamily 1408.0351}}].

\bibitem{Pogosian:2018vfr}
L.~Pogosian and A.~Zucca, \emph{{Searching for Primordial Magnetic Fields with
  CMB B-modes}}, \href{https://doi.org/10.1088/1361-6382/aac398}{\emph{Class.
  Quant. Grav.} {\bfseries 35} (2018) 124004}
  [\href{https://arxiv.org/abs/1801.08936}{{\ttfamily 1801.08936}}].

\bibitem{Brown:2005kr}
I.~Brown and R.~Crittenden, \emph{{Non-Gaussianity from cosmic magnetic
  fields}}, \href{https://doi.org/10.1103/PhysRevD.72.063002}{\emph{Phys. Rev.}
  {\bfseries D72} (2005) 063002}
  [\href{https://arxiv.org/abs/astro-ph/0506570}{{\ttfamily
  astro-ph/0506570}}].

\bibitem{Seshadri:2009sy}
T.~Seshadri and K.~Subramanian, \emph{{CMB bispectrum from primordial magnetic
  fields on large angular scales}},
  \href{https://doi.org/10.1103/PhysRevLett.103.081303}{\emph{Phys.Rev.Lett.}
  {\bfseries 103} (2009) 081303}
  [\href{https://arxiv.org/abs/0902.4066}{{\ttfamily 0902.4066}}].

\bibitem{Caprini:2009vk}
C.~Caprini, F.~Finelli, D.~Paoletti and A.~Riotto, \emph{{The cosmic microwave
  background temperature bispectrum from scalar perturbations induced by
  primordial magnetic fields}},
  \href{https://doi.org/10.1088/1475-7516/2009/06/021}{\emph{JCAP} {\bfseries
  0906} (2009) 021} [\href{https://arxiv.org/abs/0903.1420}{{\ttfamily
  0903.1420}}].

\bibitem{Cai:2010uw}
R.-G.~Cai, B.~Hu and H.-B.~Zhang, \emph{{Acoustic signatures in the Cosmic
  Microwave Background bispectrum from primordial magnetic fields}},
  \href{https://doi.org/10.1088/1475-7516/2010/08/025}{\emph{JCAP} {\bfseries
  1008} (2010) 025} [\href{https://arxiv.org/abs/1006.2985}{{\ttfamily
  1006.2985}}].

\bibitem{Trivedi:2010gi}
P.~Trivedi, K.~Subramanian and T.R.~Seshadri, \emph{{Primordial Magnetic Field
  Limits from Cosmic Microwave Background Bispectrum of Magnetic Passive Scalar
  Modes}}, \href{https://doi.org/10.1103/PhysRevD.82.123006}{\emph{Phys. Rev.}
  {\bfseries D82} (2010) 123006}
  [\href{https://arxiv.org/abs/1009.2724}{{\ttfamily 1009.2724}}].

\bibitem{Brown:2010jd}
I.A.~Brown, \emph{{Intrinsic Bispectra of Cosmic Magnetic Fields}},
  \href{https://doi.org/10.1088/0004-637X/733/2/83}{\emph{Astrophys. J.}
  {\bfseries 733} (2011) 83} [\href{https://arxiv.org/abs/1012.2892}{{\ttfamily
  1012.2892}}].

\bibitem{Shiraishi:2010yk}
M.~Shiraishi, D.~Nitta, S.~Yokoyama, K.~Ichiki and K.~Takahashi, \emph{{Cosmic
  microwave background bispectrum of vector modes induced from primordial
  magnetic fields}}, \href{https://doi.org/10.1103/PhysRevD.83.029901,
  10.1103/PhysRevD.82.121302}{\emph{Phys. Rev.} {\bfseries D82} (2010) 121302}
  [\href{https://arxiv.org/abs/1009.3632}{{\ttfamily 1009.3632}}].

\bibitem{Shiraishi:2011dh}
M.~Shiraishi, D.~Nitta, S.~Yokoyama, K.~Ichiki and K.~Takahashi, \emph{{Cosmic
  microwave background bispectrum of tensor passive modes induced from
  primordial magnetic fields}},
  \href{https://doi.org/10.1103/PhysRevD.83.123003}{\emph{Phys. Rev.}
  {\bfseries D83} (2011) 123003}
  [\href{https://arxiv.org/abs/1103.4103}{{\ttfamily 1103.4103}}].

\bibitem{Trivedi:2011vt}
P.~Trivedi, T.R.~Seshadri and K.~Subramanian, \emph{{Cosmic Microwave
  Background Trispectrum and Primordial Magnetic Field Limits}},
  \href{https://doi.org/10.1103/PhysRevLett.108.231301}{\emph{Phys. Rev. Lett.}
  {\bfseries 108} (2012) 231301}
  [\href{https://arxiv.org/abs/1111.0744}{{\ttfamily 1111.0744}}].

\bibitem{Shiraishi:2013wua}
M.~Shiraishi and T.~Sekiguchi, \emph{{First observational constraints on tensor
  non-Gaussianity sourced by primordial magnetic fields from cosmic microwave
  background}}, \href{https://doi.org/10.1103/PhysRevD.90.103002}{\emph{Phys.
  Rev.} {\bfseries D90} (2014) 103002}
  [\href{https://arxiv.org/abs/1304.7277}{{\ttfamily 1304.7277}}].

\bibitem{Trivedi:2013wqa}
P.~Trivedi, K.~Subramanian and T.R.~Seshadri, \emph{{Primordial magnetic field
  limits from the CMB trispectrum: Scalar modes and Planck constraints}},
  \href{https://doi.org/10.1103/PhysRevD.89.043523}{\emph{Phys. Rev.}
  {\bfseries D89} (2014) 043523}
  [\href{https://arxiv.org/abs/1312.5308}{{\ttfamily 1312.5308}}].

\bibitem{Durrer:2003ja}
R.~Durrer and C.~Caprini, \emph{{Primordial magnetic fields and causality}},
  \href{https://doi.org/10.1088/1475-7516/2003/11/010}{\emph{JCAP} {\bfseries
  11} (2003) 010} [\href{https://arxiv.org/abs/astro-ph/0305059}{{\ttfamily
  astro-ph/0305059}}].

\bibitem{Batchelor:1959}
G.K.~{Batchelor}, \emph{{Small-scale variation of convected quantities like
  temperature in turbulent fluid. Part 1. General discussion and the case of
  small conductivity}},
  \href{https://doi.org/10.1017/S002211205900009X}{\emph{Journal of Fluid
  Mechanics} {\bfseries 5} (1959) 113}.

\bibitem{Jedamzik:2010cy}
K.~Jedamzik and G.~Sigl, \emph{{The Evolution of the Large-Scale Tail of
  Primordial Magnetic Fields}},
  \href{https://doi.org/10.1103/PhysRevD.83.103005}{\emph{Phys.Rev.} {\bfseries
  D83} (2011) 103005} [\href{https://arxiv.org/abs/1012.4794}{{\ttfamily
  1012.4794}}].

\bibitem{Banerjee:2004df}
R.~Banerjee and K.~Jedamzik, \emph{{The Evolution of cosmic magnetic fields:
  From the very early universe, to recombination, to the present}},
  \href{https://doi.org/10.1103/PhysRevD.70.123003}{\emph{Phys.\ Rev.\ D}
  {\bfseries 70} (2004) 123003}
  [\href{https://arxiv.org/abs/astro-ph/0410032}{{\ttfamily
  astro-ph/0410032}}].

\bibitem{Dolag:99}
K.~Dolag, M.~Bartelmann and H.~Lesch{\emph{Astron. Astrophys.} {\bfseries 348}
  (1999) 351}.

\bibitem{Dolag:02}
K.~Dolag, M.~Bartelmann and H.~Lesch{\emph{Astron. Astrophys.} {\bfseries 387}
  (2002) 383}.

\bibitem{10.1007/BF02418571}
J.L.W.V.~Jensen, \emph{{Sur les fonctions convexes et les inégalités entre
  les valeurs moyennes}}, \href{https://doi.org/10.1007/BF02418571}{\emph{Acta
  Mathematica} {\bfseries 30} (1906) 175 }.

\bibitem{Zaldarriaga:1995gi}
M.~Zaldarriaga and D.D.~Harari, \emph{{Analytic approach to the polarization of
  the cosmic microwave background in flat and open universes}},
  \href{https://doi.org/10.1103/PhysRevD.52.3276}{\emph{Phys. Rev. D}
  {\bfseries 52} (1995) 3276}
  [\href{https://arxiv.org/abs/astro-ph/9504085}{{\ttfamily
  astro-ph/9504085}}].

\bibitem{Jedamzik:2020zmd}
K.~Jedamzik, L.~Pogosian and G.-B.~Zhao, \emph{{Why reducing the cosmic sound
  horizon can not fully resolve the Hubble tension}},
  \href{https://arxiv.org/abs/2010.04158}{{\ttfamily 2010.04158}}.

\bibitem{Pogosian:2020ded}
L.~Pogosian, G.-B.~Zhao and K.~Jedamzik, \emph{{Recombination-independent
  determination of the sound horizon and the Hubble constant from BAO}},
  \href{https://doi.org/10.3847/2041-8213/abc6a8}{\emph{Astrophys. J. Lett.}
  {\bfseries 904} (2020) L17}
  [\href{https://arxiv.org/abs/2009.08455}{{\ttfamily 2009.08455}}].

\bibitem{Kilo-DegreeSurvey:2023gfr}
{\scshape Kilo-Degree Survey, DES} collaboration, \emph{{DES Y3 + KiDS-1000:
  Consistent cosmology combining cosmic shear surveys}},
  \href{https://arxiv.org/abs/2305.17173}{{\ttfamily 2305.17173}}.

\bibitem{Scolnic:2017caz}
D.~Scolnic et~al., \emph{{The Complete Light-curve Sample of Spectroscopically
  Confirmed SNe Ia from Pan-STARRS1 and Cosmological Constraints from the
  Combined Pantheon Sample}},
  \href{https://doi.org/10.3847/1538-4357/aab9bb}{\emph{Astrophys. J.}
  {\bfseries 859} (2018) 101}
  [\href{https://arxiv.org/abs/1710.00845}{{\ttfamily 1710.00845}}].

\bibitem{Abbott:2017wau}
{\scshape DES} collaboration, \emph{{Dark Energy Survey year 1 results:
  Cosmological constraints from galaxy clustering and weak lensing}},
  \href{https://doi.org/10.1103/PhysRevD.98.043526}{\emph{Phys. Rev.}
  {\bfseries D98} (2018) 043526}
  [\href{https://arxiv.org/abs/1708.01530}{{\ttfamily 1708.01530}}].

\bibitem{Choi:2020ccd}
{\scshape ACT} collaboration, \emph{{The Atacama Cosmology Telescope: a
  measurement of the Cosmic Microwave Background power spectra at 98 and 150
  GHz}}, \href{https://doi.org/10.1088/1475-7516/2020/12/045}{\emph{JCAP}
  {\bfseries 12} (2020) 045}
  [\href{https://arxiv.org/abs/2007.07289}{{\ttfamily 2007.07289}}].

\bibitem{Dutcher:2021vtw}
{\scshape SPT-3G} collaboration, \emph{{Measurements of the E-Mode Polarization
  and Temperature-E-Mode Correlation of the CMB from SPT-3G 2018 Data}},
  \href{https://arxiv.org/abs/2101.01684}{{\ttfamily 2101.01684}}.

\end{thebibliography}\endgroup


\end{document}